\documentclass[preprint, 12pt]{aastex}

\usepackage{emulateapj5}
\usepackage{psfig}

\def\arcsec{$\,^{\prime\prime}$~}

\def\erg/cm2sec{ergs~cm$^{-2}$~s$^{-1}$}  
\def\ergcm2{ergs~cm$^{-2}$}  
  
\def\mdot{$\dot{m}$~}

\newcommand{\lsim }{{\lower0.8ex\hbox{$\buildrel <\over\sim$}}}
\newcommand{\gsim }{{\lower0.8ex\hbox{$\buildrel >\over\sim$}}}

\def\apj{ ApJ}
\def\aap{ A\&A}

\def\aj{AJ}

\def\Chandra{${\it Chandra}$}
\def\HST{${\it HST}$}

\def\simge{\mathrel{%
   \rlap{\raise 0.511ex \hbox{$>$}}{\lower 0.511ex \hbox{$\sim$}}}}
\def\simle{\mathrel{
   \rlap{\raise 0.511ex \hbox{$<$}}{\lower 0.511ex \hbox{$\sim$}}}}

\newcommand{\Msun}{\ifmmode {M_{\odot}}\else${M_{\odot}}$\fi}
\newcommand{\Lsun}{\ifmmode {L_{\odot}}\else${L_{\odot}}$\fi}
\newcommand{\Rsun}{\ifmmode {R_{\odot}}\else${R_{\odot}}$\fi}

\shorttitle{qLMXBs in Globular Clusters}
\shortauthors{Heinke et al.}

\begin{document}
\title{Analysis of the Quiescent Low-Mass  X-ray Binary Population in
Galactic Globular Clusters}   

\author{C. O. Heinke\altaffilmark{1}, J. E. Grindlay\altaffilmark{1},
P. M. Lugger\altaffilmark{2}, H. N. Cohn\altaffilmark{2}, 
P. D. Edmonds\altaffilmark{1}, D. A. Lloyd\altaffilmark{1}, and 
A. M. Cool\altaffilmark{3}} 

\altaffiltext{1}{Harvard-Smithsonian Center for Astrophysics,
60 Garden Street, Cambridge, MA  02138; cheinke@cfa.harvard.edu,
josh@cfa.harvard.edu, pedmonds@cfa.harvard.edu, dlloyd@cfa.harvard.edu}

\altaffiltext{2}{Department of Astronomy, Indiana University, Swain West 319,
Bloomington, IN 47405; lugger@indiana.edu, cohn@indiana.edu}

\altaffiltext{3}{San Francisco State U., 1600 Holloway Ave. San
Francisco, CA 94132; cool@quark.sfsu.edu}

\slugcomment{Submitted to ApJ}

\begin{abstract}

Quiescent low-mass X-ray binaries (qLMXBs) containing neutron stars
have been identified in several globular clusters using  \Chandra
or XMM X-ray observations, using their soft thermal spectra.   We
report a complete census of the qLMXB 
population in these clusters, identifying three additional probable
qLMXBs in NGC 6440.  We conduct several analyses of the
qLMXB 
population, and compare it with the harder, primarily CV, population
of low-luminosity X-ray sources with $10^{31}<L_X<10^{32.5}$ 
ergs s$^{-1}$.   The radial distribution of our qLMXB sample suggests an
average system mass of 1.5$^{+0.3}_{-0.2}$ \Msun, consistent with a
neutron star 
and low-mass companion.  Spectral analysis reveals that no globular
cluster qLMXBs, other than the transient in NGC  
6440, require an additional hard power-law component as often observed in
field qLMXBs.  We identify an empirical lower luminosity limit of
$10^{32}$ ergs s$^{-1}$ among globular cluster qLMXBs.   The
bolometric luminosity range of qLMXBs implies (in the deep crustal heating
model of Brown and collaborators) low time-averaged mass transfer
rates, below the disk stability criterion.  
The X-ray luminosity functions of the CV populations alone in
NGC 6397 and 47 Tuc are shown to differ.  The distribution of qLMXBs
among globular 
clusters is consistent with their dynamical formation by either
tidal capture or exchange encounters, allowing us to estimate that
seven times more qLMXBs than bright LMXBs reside in globular clusters.
The distribution of harder sources (primarily CVs) has 
a weaker dependence upon density than that of the qLMXBs.  Finally, we discuss
possible effects of core collapse and globular cluster
destruction upon X-ray source populations.   

\end{abstract}


\maketitle

\section{Introduction}

Globular clusters have proven to be an excellent place to study accreting 
binary systems, due to their known distances, ages, and reddening which 
allow system parameters and histories to be better understood than in 
the field.  It has long been suspected that globular clusters may also
 provide unique channels for the formation of accreting binaries, 
starting with the discovery that low-mass X-ray binaries (LMXBs) are 
$\sim100\times$ more common
(per unit mass) in globular clusters than in the field (Clark 1975;
Katz 1975).
It is now thought that X-ray binary systems in dense globular clusters
are created principally through exchange interactions between
primordial binaries and other stars (see, e.g. Hut, Murphy \& Verbunt 1991).
Therefore, studying accreting binary systems in globular   
clusters can give us insight into the characteristics of accreting
binary systems and populations, and also into dynamical 
effects inside globular clusters.
 
The giant leap forward in sensitivity and resolution offered by the 
\Chandra\ X-ray Observatory, especially when combined with the 
unparalleled resolution of the {\it Hubble Space Telescope} (\HST), has 
revolutionized our understanding of globular 
clusters. Prior to \Chandra, there were 12 bright cluster LMXBs (now 13,
all thought to have neutron stars [NS] as primaries), 
and 57 faint X-ray sources known in the Galactic globular
cluster 
system (Verbunt 2001).  A few of the latter had been identified 
with cataclysmic variables (CVs; Hertz \& Grindlay 1983, Cool et
al. 1995, Grindlay et al. 1995)
, and some were thought to be LMXBs in quiescence (qLMXBs; Verbunt,
Elson \& van Paradijs  
1984), but their properties were poorly understood due to the poor 
spatial and/or spectral resolution of previous X-ray observatories.  
\Chandra\ has identified more than 100 X-ray sources in the 
globular cluster 47 Tuc alone (Grindlay et al. 2001a, hereafter
GHE01a).  Dozens of 
sources have been discovered in the globular clusters $\omega$ Cen
(Rutledge et al. 2002a, 
 Cool, Haggard, \& Carlin 2002), NGC 6397 (Grindlay et al. 2001b,
hereafter GHE01b), NGC 6752 (Pooley 
et al. 2002a), NGC 6440 (Pooley et al. 2002b; in't Zand et al. 2001),
M28 (Becker et al. 2003), Terzan 5 (Heinke et al. 2003b), and M80 (Heinke
et al. 2003c).  These plus preliminary results from several additional clusters
 create a sizable dataset for comparison of X-ray source populations with
cluster properties (Pooley et al. 2003a, b), especially when combined
with initial XMM results on the low-density clusters M22, $\omega$
Cen, and M13 (Webb, Gendre \& Barret 2002; Gendre et al. 2003a, b). 
  \HST\ identifications have  
allowed the classification of many as CVs or active main-sequence or 
subgiant binaries (ABs) in 47 Tuc (Edmonds et al. 2003a,b), NGC 6397
(GHE01b), NGC 6752 (Pooley 
et al. 2002a), and $\omega$ Cen (Cool, Haggard, \& Carlin 2002), and
one as a qLMXB in 47 Tuc (Edmonds et al. 2002a).  Radio-derived
positions or orbital periods 
for millisecond pulsars (MSPs) have allowed the identification 
 of faint X-ray sources as MSPs in 47 Tuc (Edmonds et al. 2001;
Grindlay et al. 2002; Edmonds et al. 2002b), 
NGC 6397 (GHE01b), and NGC 6752 (D'Amico et al. 2002).  However, radio and
\HST\ coverage of globular clusters is generally quite incomplete, and faint 
X-ray sources in most globular clusters do not have clear
identifications from other wavebands.   
Even with deep radio and optical data, MSPs may be confused with CVs (see 
Edmonds et al. 2002b) or with ABs (Ferraro et al. 2001,
but cf. Edmonds et al. 2003b).  However, qLMXBs show clear differences
from all  
other types of globular cluster X-ray sources in their colors and 
luminosities ($L_X\sim10^{32-34}$ ergs s$^{-1}$, with soft X-ray 
colors) and detailed spectra (a thermal component with implied radius
of a few km, sometimes accompanied by a nonthermal harder component).
Their X-ray emission is thought to be due principally to thermal
emission from the neutron star surface (Brown, Bildsten, \& Rutledge
1998; Campana et al. 1998), thus making their X-ray 
signatures more homogeneous than X-rays from other source types.  
  These differences make it practical to identify a homogeneous sample 
of qLMXBs in different globular clusters without the need for deep \HST\ 
or radio datasets, using X-ray spectra, colors, and luminosities 
alone, as for instance in 47 Tuc (GHE01a), $\omega$ Cen (Rutledge et
al. 2002a), NGC 6397 (GHE01b), NGC 6440 (Pooley et al. 2002b), and 
M28 (Becker et al. 2003).

We perform several analyses of the qLMXB population that has been 
uncovered in numerous globular clusters.  The qLMXB population in 
globular clusters offers hope for understanding many questions 
related to neutron stars, accretion flows, and cluster dynamics.  
Among these questions: what is the source of the X-ray luminosity
observed from qLMXBs, deep crustal heating (Brown et al. 1998) or
continued low-level accretion?  What is the nature  
and origin of the nonthermal hard component observed in many qLMXB 
spectra?  How long are qLMXBs in quiescence between outbursts?  Are 
qLMXBs part of the progenitor population of MSPs?  Can we constrain 
the radius and/or mass of qLMXBs through observation of the thermal 
component of their spectra?  What are the parameters of globular 
clusters that cause the formation of qLMXBs? Can their numbers be 
fully explained through two-body or single-binary encounter rates?
Are other globular cluster X-ray sources (such as CVs) formed in the
same way?  We will gather evidence to begin answering these questions in this 
paper.

 In \S 2 we identify a sample of confirmed and potential
qLMXBs containing neutron stars in several globular clusters, and
analyze their X-ray spectra.  In \S 3.1 we study the qLMXB radial
distributions in King-model clusters. In \S 3.2 we analyze the qLMXB
(and harder sources) luminosity functions, while in \S 3.3 we 
analyze the dependencies of qLMXB and harder source numbers upon
cluster structural parameters.  In \S 4 we discuss the
meaning of qLMXB radial distributions and spectra (\S 4.1), qLMXB
luminosities (\S 4.2), qLMXB vs. CV distributions among clusters (\S 4.3), 
and additional dynamical processes (\S 4.4).
Finally we summarize in \S 5.

\section{Quiescent LMXBs}

Identifying the nature of X-ray sources is often difficult, requiring
deep multiwavelength data.  The unique X-ray spectral signature of a qLMXB,
however, offers 
the possibility of identifying qLMXBs without multiwavelength
followup (important since they can be extremely optically faint; see
Edmonds et al. 2002a, GHE01b).  In the field, 
qLMXBs containing neutron stars have been identified after bright
transient outbursts, often exhibiting type I X-ray bursts
confirming their neutron star nature. (See Campana et al. 1998 for a review.)
Their quiescent spectra 
show a thermal component roughly consistent with a 10 km neutron
star (when hydrogen atmospheres are taken into account, e.g. Brown
et al. 1998).  In addition, a harder component
parametrized as a power-law of 
photon index 1 to 2 is often seen, comprising up to 40\% of the 0.5-10
keV emission (Campana et al. 1998; Rutledge et al. 2002b and refs
therein).  Their  
minimum X-ray (0.5-2.5 keV) luminosity appears to range between
$5\times10^{31}$ and a few $10^{33}$ ergs s$^{-1}$, although
distances and thus luminosities are uncertain.  Comparison with field
systems has thus 
allowed numerous qLMXBs to be identified in X-ray studies of globular
clusters (see below). 
 
The large sample of qLMXBs now known in several globular clusters
allows significant comparative study.  The known distances and
reddening to globular clusters allows accurate luminosities to be
derived, generally impossible in the field.  We can fit the spectra to
look for a hard power-law component as seen in some field qLMXBs, and
to check for consistency with a 10 km, 1.4 \Msun neutron star
explanation.  Calibration changes since the publication of some early
papers makes 
revisiting the spectral analysis on several clusters desirable, while
a common standard for identifying qLMXBs would also be useful.
For all analysis in this paper, we
use photoelectric absorption X-ray cross-sections of Balucinska-Church
\& McCammon (1992) in the XSPEC {\it phabs} model.

For this paper we choose to identify a ``canonical'' qLMXB signature,
spectral consistency with a nonmagnetic hydrogen atmosphere of implied
radius $\sim$10 km, with a
small ($<40$\% of 0.5-10 keV flux) contribution from a power-law
component allowed.  This is chosen to match qLMXB systems studied 
in the field (Cen X-4, Rutledge et al. 2001; Aql
X-1, Rutledge et al. 2002b; 4U 2129+47, Nowak et al. 2002), and gives
a spectrum much softer than that from known CVs.  A notable
exception to this signature is the 
millisecond X-ray pulsar SAX J1808.4-3658 (SAX J1808), which has shown
an extremely faint
($L_X$(0.5-10 keV)$=5\times10^{31}$ erg s$^{-1}$) quiescent
spectrum
in a recent XMM observation dominated by a hard (photon index
$\sim1.5$) power-law component (Campana et al. 2002).  We would not be
able to distinguish such an object from CVs with similar spectra in
globular clusters.  Therefore we concentrate in this paper on studying
a sample of objects which we can be fairly certain are qLMXBs.
 
As a comparison sample, we discuss spectrally harder objects of similar X-ray
luminosity.  Based on optical analysis in several clusters (GHE01a, GHE01b,
Pooley et al. 2002a, Edmonds et al. 2003a), these are thought to be mostly
CVs, at least above $10^{31}$ erg s$^{-1}$ (below which active binaries and
MSPs become numerous).  One hard X-ray source at the upper end of this
luminosity range has been identified as a bright MSP in the globular
cluster M28 (Becker et al. 2003), while a moderately hard source in 47 Tuc
with an unusual spectrum and strong variability has been suggested as a
qLMXB (X10; GHE01a, Edmonds et al. 2003b).  Transient black holes in
quiescence have X-ray spectra and luminosities indistinguishable from those
of CVs, although none have yet been positively identified in a globular
cluster. The great majority of these hard objects are probably CVs, as
we show below.  

CVs, composed of a (usually low-mass
main-sequence) secondary, a white dwarf, and an accretion disk, display
several optical signatures. In $U$ vs. $U-V$ CMDs they appear bluer than
the main sequence, usually lying between the main sequence and the white
dwarf cooling sequence. A strong contributor to this blue color, the
accretion disk, will also generate H-$\alpha$ emission, and cause
short-timescale nonperiodic variability (flickering) and sometimes
large-amplitude outbursts. In $V$ vs. $V-I$ CMDs, CVs generally appear
redder due to the increased contribution of the secondary light, and in
globular clusters have often been observed to fall on or near the main
sequence (Edmonds et al 2003a).  The secondary, filling its Roche lobe,
will often show ellipsoidal variations, and this periodic, low amplitude
signal is detectable if the noise from flickering is not too large. The
X-ray to optical flux ratio for CVs should be smaller than for qLMXBs, and
a CV at a given X-ray luminosity will display much bluer colors than a
qLMXB at the same luminosity, since it must be accreting at a much higher
rate. Based on these characteristics, X-ray sources can be identified as
CVs with high confidence.

Detailed searches for variability and/or H-$\alpha$ excess among blue
counterparts to \Chandra\ X-ray sources have been published for three
clusters, NGC 6397, NGC 6752, and 47 Tuc.  Six CVs have been optically
identified among 8 hard X-ray sources with $L_X(0.5-2.5)>10^{31}$ ergs
s$^{-1}$ in NGC 6397 (GHE01b). The two unidentified objects are an active
binary system and a probable CV (based on an $N_H$ column above the cluster
value). Similarly, six CVs have been optically identified in NGC 6752
(Pooley et al. 2002a) among 8 hard cluster sources with
$L_X(0.5-2.5)>10^{31}$ ergs s$^{-1}$, including one probable BY Dra system
and one unidentified source. (See \S 2.5 for discussion of the soft source
CX8 in NGC 6752.)  Finally, 12 CVs have been identified among 18 hard
sources with $L_X(0.5-2.5)>10^{31}$ ergs s$^{-1}$ in 47 Tuc (GHE01a,
Edmonds et al. 2003a, b). The other six include three active binary systems
and three sources with marginal optical counterparts that are suggested as
CVs.  Thus the fraction of non-CVs among this population seems to be less
than $\sim$20\% in three very different clusters.

Quiescent LMXBs have been spectrally identified in 47 Tuc (2; GHE01a),
$\omega$ Cen 
(1; Rutledge et al. 2002a), NGC 6397 (1; GHE01b), NGC 6440 (4-5; Pooley et
al. 2002b, in't Zand et al. 2001), M28 (1; Becker et al. 2003), Terzan
5 (4 plus the transient LMXB; Heinke et al. 2003b), 
M13 (1; Gendre et al. 2003b), M30 (1; Lugger et al. 2003),
M80 (2; Heinke et al. 2003c), and NGC 6266 (5; Pooley et al. in
prep.).  Spectral  
analyses have been performed for most of these.  However, in several cases
the work was performed before the low-energy \Chandra\ quantum efficiency
degradation was known and calibrated, so we repeat the analysis.  We
also attempt a census of all qLMXBs in clusters studied with \Chandra\ or
XMM.  We summarize some important quantities, and results from 
\Chandra\ or XMM studies of several clusters, in Table 1 (with clusters
listed in order of decreasing close encounter rates, see \S 3.3 and \S
4.3).  Physical
quantities are from the catalog of Harris 1996, updated
2003\footnote{Available at
http://physun.physics.mcmaster.ca/Globular.html}, with a few
additional updates, principally core radii and distances (referenced).

\subsection{47 Tuc \& M30}

These clusters have been recently analyzed (Heinke et al. 2003a,
Lugger et al. 2003) with the same nonmagnetic
hydrogen-atmosphere model (Lloyd 2003) used in this work.  We include
the relevant results from their analyses of these qLMXBs in Table
2. We correct the luminosities and radii for the fraction of the
1 keV \Chandra\ point spread function included in the extraction circles,
using the CIAO tool {\it mkpsf}.  The
luminosity corrections amount to factors of 1.05 and 1.10 for 47 Tuc
and M30 respectively; this small correction was not applied to the 47 Tuc
results in Heinke et al.~(2003a).  
The possibility that X10 in 47 Tuc may be a qLMXB was suggested by
Edmonds et al. (2003b) on the basis of its high X-ray to optical
flux ratio.  However, this object is dramatically variable in the
X-ray (GHE01a), 
while the X-ray and optical flux measurements were not simultaneous
(as pointed out by Edmonds et al. 2003b).
Its X-ray spectrum can be described as a power-law of photon index 3, with no
apparent contribution from a thermal hydrogen atmosphere component
(GHE01a).  We also note that the optical counterpart V3, while on the main
sequence in a $V$, $V-I$ CMD, falls near the white dwarf cooling tracks 2.8
magnitudes bluewards of the main sequence in a
$U$, $U-V$ CMD (Edmonds et al. 2003a).  This indicates a very strong
contribution of U light from the disk, similar to several other CVs
apparent in the 47 Tuc CMDs of Edmonds et al. (2003a). This can be
compared to the 
known qLMXB X5, which falls roughly 0.9 magnitudes bluewards of the
main sequence in a F300W, F300W-F555W CMD (Edmonds et al. 2003a).  The
difference in $U$ contributions suggests that significantly more mass
transfer is occurring in X10 than X5.  X5's high inclination could
reduce its disk component, but the eclipsing CVs W8 and W15 also
appear very blue (Edmonds et al. 2003a).  
Further investigation of the X-ray and optical properties of this
object in the coordinated simultaneous \Chandra/\HST\ observations of 47 Tuc in
late 2002 is underway, and should resolve the question of X10's nature. 
 For the purposes of this paper, we simply note that X10 does not fit
our ``canonical'' qLMXB definition above, and so we exclude it from our
qLMXB analysis.

\subsection{M28 \& M13}

Quiescent LMXBs in these clusters have been recently analyzed using
the Zavlin, Pavlov, \& Shibanov (1996) nonmagnetic hydrogen atmosphere
{\it nsa} model (Becker et al. 2003, Gendre et al. 2003b).  This model
gives very similar results to the Lloyd (2003) model.  Since the data
are not yet public, we include the results of their analyses in our
table.  Since we are quoting true radii and T$_{eff}$ rather than
R$^{\infty}$ and T$^{\infty}_{eff}$, we calculate the temperature and
radius results using the assumption of z=0.306.  The luminosity values
given in Gendre et al. (2003b) and Becker et al. (2002) are not for 
the same energy ranges as our 0.5-2.5 keV range.  We used the Zavlin
et al. (1996) 
{\it nsa} model, with the same temperatures and radii as the best fits
in these papers, to
calculate the ratios of unabsorbed flux in our energy bands to their energy 
bands, and thus generated the luminosities in Table 2. 

\subsection{$\omega$ Cen, NGC 6397, Terzan 5, and M80}

For these clusters we re-extract source and background spectra of
qLMXB candidates in the 
same way as in the original works (Rutledge et al. 2002a, GHE01b,
Heinke et al. 2003b, Heinke et al. 2003c), except that we use an extraction
circle of 2'' for NGC 6397 U24.  We calculate 
the fraction of the 1 keV \Chandra\ point-spread function that is included in
each region using the CIAO tool mkpsf.   We applied a correction to
the effective area functions to account for degradation of low-energy
quantum efficiency. We group the channels to ensure
$\geq$20 counts per bin, and fit the data in XSPEC (Arnaud 1996).  For
M80 CX6 (only 62 counts), we do not group the channels, 
and instead use the C statistic for analysis.  This would be
preferable for the faint sources in Terzan 5, but the background is
extremely high due to the outburst by the transient LMXB.  When we use the C 
statistic, we test the fit by generating Monte Carlo simulations of
the best fit spectrum.  Roughly half the Monte Carlo simulations
should show lower values of the C statistic than the data, if the fit
is good.  The percentage of simulations showing a lower C statistic
than the best fit is shown in Table 2 for sources where we use the
C statistic.

For each qLMXB we fit a model consisting of photoelectric absorption
fixed at the cluster value and the nonmagnetic hydrogen atmosphere of
Lloyd (2003), plus an optional power-law component.   The power-law
component in field systems has been observed with a photon index
ranging from 1 to 2, and contributing up to 40\% of the flux (see
Rutledge et al. 2002b; but cf. Campana et al. 2002 on SAX J1808).  To
simplify our analysis and allow direct comparison between objects, we
fix the photon  
index of the possible power-law component at 1.5.  The results of these
fits are shown in Table 2.  Luminosities are increased to account for
aperture corrections by factors of 1.10 for Terzan 5, 1.02 for
$\omega$ Cen, 1.03 for NGC 6397, and 1.07 for M80; derived radii are adjusted
accordingly.   All spectral errors, but not luminosity
errors, are given at 90\% confidence for a single parameter using the
XSPEC {\it error} command.  For
power-law component upper limits, this translates to 95\% confidence.  We note
that this procedure may significantly underestimate strongly covariant
uncertainties, such as radius and temperature, especially when the
absorption is large (see in't Zand et
al. 2001).   Therefore we offer radius estimates as a consistency
check, and not as a serious attempt as a constraint upon the radii of these
objects, especially for small numbers of counts.  This is especially
true since we have not considered intrinsic absorption in these
systems.  The values for Terzan 5 are especially questionable due to
the extremely high background; for W4 we make no attempt to constrain
any power-law component (although none is required).  We include these
parameters for completeness and to generate luminosities.  Upcoming
\Chandra\  observations of Terzan 5 in July 2003 (principal
investigator: R. Wijnands) should allow 
much better constraints upon these sources if the transient remains
in quiescence as expected.

\subsection{NGC 6440}

For NGC 6440 we downloaded the data from the archive and removed the
pixel randomization applied in 
standard processing.  Then we extracted spectra from 
1\farcs2 radius circles around {\it wavdetect}-identified locations of
the softer sources (CX1, CX2, CX3, CX5, CX7, CX8, CX10, CX11, CX12, CX13) from 
Pooley et al. (2002b), except where severe crowding reduced our region
size (to 0\farcs86 for CX11).  We grouped counts to 20 per bin
to ensure $\chi^2$ statistic applicability except when the source had
fewer than 100 counts,
where we left the data unbinned and used the C statistic in XSPEC for
analysis (as above).   We fix the absorption to the cluster value of
0.59$\times10^{22}$ 
cm$^{-2}$ (as in Pooley et al. 2002b) for all sources.  
We demand that candidate qLMXBs be acceptably fit with a hydrogen
atmosphere alone, or require that the power-law component comprise no more
than 40\% of the 0.5-10 keV flux (see above). All luminosities and radii are
adjusted to account for the extraction regions; this amounts to a
factor of 1.07 in luminosities for the qLMXBs we identify. 
In this way, we find several additional qLMXB candidates in NGC 6440.

The quiescent spectrum of the X-ray transient NGC 6440 CX1 was
analyzed by in't Zand et 
al. (2001), who assume a higher $N_H$ column (0.82$\times10^{22}$
cm$^{-2}$) than the galactic column to the cluster.  They base this
upon the agreement between the $N_H$ columns derived by a 1998 BeppoSAX
observation of CX1 in outburst (in't Zand et al. 1999), and their
analysis of the spectrum of an annulus around the outbursting
transient in a 2001 \Chandra\ observation.  
However, the BeppoSAX spectrum gives different $N_H$ values
depending on the assumed model, and the \Chandra\ 2001 annulus spectrum
of the outburst will be affected by the energy dependence of the
point-spread function.  Also, internal absorption is often observed to vary
in accreting systems, especially during outbursts.   Therefore,
we perform our own determination of $N_H$ for 
the quiescent spectrum of CX1.  When we allow $N_H$ to float, we find 
$N_{H,22}=0.66^{+.11}_{-.10}$.  Using the {\it nsa} model of Zavlin et
al. (1996) and forcing the power-law photon index to be 1.44 (as done
by in't Zand et al. 2001), but using more recent photoelectric
cross-sections, we find $N_{H,22}=0.70^{+.18}_{-.10}$ (and
R=9.5$^{+0.7}_{-1.5}$ km). We consider the
absorption to CX1 to remain uncertain, and quote results assuming the
cluster absorption in Table 2.

Our results strongly support the suggestion by Pooley et al. (2002b)
that CX2, CX3, CX5, and CX7 are likely qLMXBs.  Although the best fits
for CX3 and CX5 include a power-law component, this component cannot be
considered significantly detected. We also identify three additional possible
qLMXBs, CX10, CX12, and CX13.  Each of these objects is consistent
with a 10 km hydrogen-atmosphere neutron star spectrum, without a
power-law component.  Their luminosities are also similar to those 
of known qLMXBs in $\omega$ Cen (Rutledge et al. 2002a), NGC 6397
(GHE01b), and M80 (CX6; Heinke et al. 2003c).  None of these objects
show clear variability, though the number of counts from each is low.
We include their
spectral parameters in Table 2, although we note that their low counts
make these parameters individually unreliable.

For the harder objects CX8 and CX11, we can rule out a hydrogen
atmosphere as the primary contribution to the observed flux.  
For CX8 a 10-km radius hydrogen atmosphere is a bad fit (100 Monte Carlo
simulations of the best fit gave uniformly
lower values of the C statistic than the data). A hydrogen atmosphere with an
additional power-law component is an acceptable fit, but has
70$^{+30}_{-24}$\% of 
the unabsorbed 0.5-10 keV flux (92\% of the received flux) in the
power-law component.
For CX11 a 10 km-radius hydrogen atmosphere alone is as bad a fit as
for CX8.  A hydrogen atmosphere with an
additional power-law component is an acceptable fit, but has
47$^{+30}_{-24}$\% of 
the unabsorbed 0.5-10 keV flux (83\% of the received flux) in the
power-law component.   Although the uncertainties of the fit allow
less than 40\% of the unabsorbed flux to be in the power-law
component, we think this object is unlikely to be a qLMXB.  The
remaining sources above CX11's luminosity are too hard to fit our
canonical qLMXB model.  

Fainter sources in the observation are uniformly harder than our
suggested qLMXBs, and have too few counts for even the
simplest spectral fitting to be meaningful.  We note that CX13 is as
faint (or fainter) as the faintest other qLMXBs identified in any
globular cluster. 
Although confusion may have prevented the identification of perhaps
one qLMXB, we conclude that this census of the qLMXBs in NGC 6440 is
essentially complete.  Follow-up observations of NGC 6440 scheduled
for July 2003 (principal investigator: R. Wijnands) will allow testing
of these results.

Finally, we utilize a simple graphical method to check the similarity of
our candidate qLMXBs to other qLMXBs.  We place each of the NGC 6440 qLMXB
candidates on a standardized X-ray CMD, using our best-fit
luminosities and the unreddened colors derived by Pooley et
al. (2002b).  We also calculate the positions of several other qLMXBs
in other clusters on the same diagram.  To do this requires
compensating for the reddening to different clusters, as well as
accounting for the differences between the response of the
front-illuminated ACIS-I chips (used for the observations of 47 Tuc,
NGC 6397, $\omega$ Cen, and Terzan 5) and the back-illuminated ACIS-S chip.
We used the CIAO PIMMS tool to estimate the difference in observed
Xcolor for several spectra between each cluster's actual observation
and a hypothetical 
ACIS-S observation with no reddening (ignoring the loss of low-energy
sensitivity over the \Chandra\ mission, which has no 
effect on the spectra above 1 keV).  
We derive Xcolor offsets of +0.70 for $\omega$ Cen, +0.49 for 47 Tuc, +0.73
for NGC 6397, and +0.06 for M30 to place their qLMXBs onto a dereddened
ACIS-S CMD.  The two qLMXBs in 47 Tuc suffer moderate pileup, which
distorts their intrinsic Xcolor.  To reduce the hardening effect of
pileup, we use the Xcolor derived only from the final 2000 47 Tuc
observation (4.7 ksec), which used a 1/4 subarray 
to avoid pileup (receiving 369 and 423 counts for X5 and X7
respectively).  However, we note that these objects (and the M30
qLMXB) still suffer pileup at the $\sim2$\% level, which will harden
their spectra and is not accounted for. Finally, we plot the
theoretical locations in this 
diagram of hydrogen atmosphere neutron stars of 10 and 12 km radii
over a range of temperatures, using the models of Lloyd (2003) with
the gravitational redshift fixed to z=0.306.  These
are essentially neutron star cooling tracks.

This X-ray CMD (Figure 1) clearly shows agreement between the
theoretical hydrogen 
atmosphere cooling tracks and the observed locations of identified and
candidate qLMXBs.  We caution that the apparent close agreement with the
12-km track should not be taken too seriously due to the numerous
possible errors listed above.  We 
also plot the theoretical track of a blackbody, 
with radius arbitrarily set to 1.41 km.  A clear difference can be
seen between the predictions of a blackbody vs. the hydrogen
atmosphere models, with the colors and luminosities of the qLMXBs favoring the
hydrogen atmosphere model.  This diagram shows the utility of a 
standardized X-ray CMD, and the importance of hydrogen-atmosphere
neutron star models.  

\subsection{Other clusters}

Several other clusters have been studied with \mbox{\Chandra's} ACIS
detector to a 
depth sufficient to identify any of the qLMXBs above.  Pooley et
al. (2002a) find no objects in NGC 6752 with luminosities and spectra
similar to those found in other clusters.  However, they do identify a
source (CX8) within the 
half-mass radius of NGC 6752 which shows an extremely soft spectrum, placing it
near the neutron star cooling track in Figure 1, at an implied 
$L_X=2\times10^{31}$ ergs/s.  We note that no thermally radiating MSP
has been identified at these luminosities; from Grindlay et al. (2002),
we see that the 0.5-2.5 keV X-ray luminosities of thermally emitting
MSPs range from 1-4$\times10^{30}$ ergs/s.  Nonthermally-emitting MSPs
can be brighter (GHE01b, Edmonds et al. 2002b, Becker et al. 2003),
but  also have much harder spectra.  Identification of a thermally
emitting neutron star in the ``gap'' between 
the tail of the qLMXB luminosity function and the MSP nexus would be
of great interest.

We extracted the spectrum of CX8 from the \Chandra\ archival
observation of NGC 6752, using a 2\arcsec extraction region and binning
the 83 counts with 10 counts/bin.  XSPEC fits with a hydrogen atmosphere
model give atrocious fits; fixing the radius to 10 km and the
absorption at the cluster value gives
$\chi^2_{\nu}$=7.4 for 8 degrees of freedom (dof), while allowing the
radius to be a free 
parameter (R$\sim0.9$ km) gives $\chi^2_{\nu}$=2.4 for 7 dof.
The poor quality of these fits is induced by a feature resembling an
 emission line 
complex located at $\sim0.9$ keV.  This indicates an optically
thin low-temperature plasma, so we fit an XSPEC MEKAL optically thin
plasma model (Liedahl, 
Osterheld, \& Goldstein 1995 and refs therein), deriving
$kT=0.77\pm.11$ keV and an iron abundance $23^{+38}_{-12}$\% of
solar.  This iron abundance is inconsistent with the metallicity of
NGC 6752, [Fe/H]=-1.65 (3\% solar), from Harris (1996).  The
metallicity inconsistency and 
the large ($\sim88$\arcsec, 8.5 core radii) offset of CX8 from the core of
NGC 6752 suggest that CX8 may not be a member of the cluster, but
rather a foreground star or active binary.  

Preliminary analyses of deep \Chandra\ or XMM data have been presented
for five other clusters at 
scientific conferences or in other works at the time of writing.
According to both Pooley et al. (2003a) and Gendre et al. (2003b), NGC
6366 possesses no qLMXBs. NGC 6121 has no qLMXBs (Bassa et al. in
prep.; Pooley et al. 2003a), nor does NGC 5904 (Pooley et
al. 2003a). The rich, dense cluster NGC 6266 (Pooley et al. 2003b,
Pooley et al. in prep.) has five identified qLMXBs (D. Pooley,
priv. comm.).  Reassessment 
of the XMM data on M22 (Webb et al. 2002) by Gendre et al. (2003b)
indicates that no qLMXBs matching our template are present in that
cluster. In each case 
the sensitivity of the observation is sufficient to confidently identify all
possible qLMXBs above our empirical lower $L_X$ limit of
$1\times10^{32}$ erg s$^{-1}$.  In most of these clusters the
sensitivity is sufficient to identify all harder
sources above $L_X$(0.5-2.5 keV)$>10^{31}$ ergs s$^{-1}$ (Pooley et
al. 2003a, b; see Table 1).
We use these preliminary results in 
our analysis of cluster weightings in \S 3.3.

\section{Comparative Analysis}

\subsection{qLMXB Spatial Distribution}

\citet{grindlay84} estimated the typical mass of the bright LMXBs in
eight globular clusters by analyzing the distribution of the radial
offsets of these sources relative to the cluster centers.  This
approach assumes that the sources have a common mass and that the
cluster potentials have a common structure.  It is also assumed that
the distributions of X-ray sources and normal stars are in thermal
equilibrium.  In this case, the most massive group of normal stars is
expected to be approximately distributed as a \citet{king66} model.  This
analysis results in an estimate for the ratio $q = M_X/M_\ast$ of the
source mass to the mass of the typical star that defines the
optical core radius.  
Using a maximum likelihood analysis, \citet{grindlay84} found a most
likely value of $q=2.6$ with a 90\% confidence range of $1.8-3.8$.  In
this section, we adapt this analysis to the qLMXB distribution, in
order to estimate the qLMXB mass.

\citet{grindlay02} describe a maximum-likelihood appoach for
fitting ``generalized King models'' to the projected radial
distributions of cluster objects.  In this model, the projected
surface density of each component is described by,
\begin{equation}\label{gen_king_model}
S(r) = S_0 \, \left[1 + \left({r \over r_0 }\right)^2 \right]^{\alpha/2},
\end{equation}
where $\alpha$ is the power-law index and the core radius $r_c$ is
related to the radial-scale parameter $r_0$ by $r_c = (2^{-2/\alpha} -
1)^{1/2}\,r_0$.  \citet{grindlay02} obtained independent
estimates of $r_c$ and $\alpha$ for various source populations in
47~Tuc by maximum likelihood fits of Eqn.~(\ref{gen_king_model}).  If
the optical core radius $r_{c\ast}$ of the cluster is defined by stars
of mass $M_\ast$ that have a standard King-model distribution ($\alpha_\ast
= -2$), then the core radius and slope for the distribution of sources
of mass $M_X = q M_\ast$ are related to $q$ by,
\begin{mathletters}
\label{rc_alpha}
\begin{eqnarray} 
r_{cX}  & = & \left(2^{2/(3q -1)} - 1 \right)^{1/2} r_{c\ast}  \\
\alpha_X & = & 1 - 3q 
\end{eqnarray}
\end{mathletters}

In the present study, we first explored two-parameter
maximum-likelihood fits of Eqn.~(\ref{gen_king_model}) to the
distribution of 20 qLMXBs in seven clusters with King-model structure (47
Tuc, NGC 6440, Terzan 5, M80, M28, $\omega$ Cen, and M13---note that we
include the Terzan 5 transient LMXB).  However, this produced
large uncertainty ranges for both $r_{cX}$ and $\alpha_X$, and thus did
not provide much useful information on the qLMXB mass.  We then
performed single-parameter fits by determining the value of $q$ that
maximizes the likelihood of observing the given sample, with $r_{cX}$
and $\alpha_X$ given by Eqn.~(\ref{rc_alpha}).  Thus, we fit the
function,
\begin{equation}\label{one_par_fit}
S(r) = S_0 \, \left[1 + \left({r \over r_{c\ast}}\right)^2 \right]^{(1-3q)/2}.
\end{equation}
This is similar to the approach of \citet{grindlay84}.  
We estimate the uncertainty range for $q$ by fitting 1000 bootstrapped
resamples of the original 20-source sample \citep[see][]{cohn02}.  The
result is $q = 1.9\pm0.2$\,(1-$\sigma$) with a 90\% confidence range
of $1.6 \le q \le 2.2$.  The corresponding values of core radius and
slope are $r_{cX}=(0.60\pm0.04)\,r_{c\ast}$ and $\alpha_X=-4.5\pm0.5$.

We note that the assumption of thermal equilibrium among the cluster
objects is not strictly justified for $\omega$ Centauri, which has not
reached equipartition (Anderson, 1997).  Therefore, we removed the
qLMXB in $\omega$ Cen from our sample and repeated the analysis, and
derived the same results.  We performed the same analysis upon
the sample of eight LMXBs used in \citet{grindlay84}, finding $q=2.2$
with 90\% confidence range of $1.7 \le q \le 3.7$.  This is in
reasonable agreement with their determination of $q=2.6$, $1.8 \le q \le
3.8$ (90\% confidence), with the major difference being that the
current analysis does not consider offset measurement errors
(negligible for \Chandra\ positions).  Finally, we performed the same
analysis upon the sample of soft X-ray sources in 47 Tuc from Grindlay
et al. (2002), differing from that analysis by parameterizing both the
core radius and power-law slope $\alpha$ in terms of a single $q$.  Our
new value for these soft sources is $q=1.58\pm0.13$, 90\%
confidence range $1.40 \leq q \leq 1.81$.  

If we assume that the optical profiles of the sample clusters, from
which the core radii were determined, are dominated by turnoff-mass
objects, then a reasonable estimate for $M_\ast$ is $0.8~\Msun$ (see,
e.g., King et al.~1998).  This
results in a most-likely qLMXB mass of $M_X=(1.5\pm0.2)~\Msun$ with a 90\%
confidence range of $1.3~\Msun \le M_X \le 1.8~\Msun$.   This range
comfortably allows for a Chandrasekhar-mass neutron star 
with a low-mass companion.  For the soft
sources in 47 Tuc (which probably include some ABs as well as MSPs, Edmonds et
al. 2003b), we derive a mass range of 
$M_X=1.26\pm0.10$ \Msun, 90\% confidence range $1.12~\Msun \leq M_X \leq
1.45~\Msun$.  For the eight LMXBs from Grindlay et al. (1984), our 90\%
confidence range is $1.4~\Msun \le M_X \le 3.0~\Msun$.  Both of these
estimates are also consistent with the qLMXB range.

\subsection{Luminosity Functions}

Recent work by Pooley and collaborators (Pooley et al. 2002b, 2003a)
has shown clear differences between luminosity functions (LFs) of
different clusters, attributed in large part to differences in source makeup
(e.g. large numbers of MSPs and ABs in 47 Tuc compared to NGC 6397).
The LFs of globular cluster X-ray sources should be
affected both by the relative numbers of different sources in each
cluster, and by the properties of the individual populations.
Here we make a first attempt to characterize the LFs for
two identifiable source populations, qLMXBs and harder sources with
$L_X>10^{31}$ ergs s$^{-1}$, identified as mostly CVs (see \S 2).  For
the latter group, we test whether the hard source luminosity functions
in different clusters are consistent, in order to identify possible
differences between the CV populations of different clusters. 
  
To study the luminosity distributions of these populations, we
follow the formalism of Johnston \& Verbunt (1996), and Pooley et
al. (2002a).  Due to the relatively small numbers of sources, we
assume for this analysis a simple power-law shape for the LF above a
limiting luminosity. We derive the  
best-fit luminosity function by forming the quantities
$z_j=(L_i^j/L_i)^{-\gamma}$ and finding the $\gamma$ that most
uniformly distributes the $z_j$ along the interval [0,1].  Here $L_i$
is the limiting luminosity to which the cluster has been searched, or
the minimum luminosity of the analysis.  If the true LF is not a
power-law, using differing limiting luminosities for different
clusters will generate apparent differences between LFs of different
clusters. Therefore we attempt to use the same limiting luminosity for
each cluster within one analysis.  
 For qLMXBs we include all qLMXBs in \S 2,
analyzing both the luminosities generated with 10 km fits and those
where the radius was allowed to float.  We take
as our minimum luminosity the lowest 
luminosity of a detected qLMXB. This leaves only Terzan 5 with a higher
limiting luminosity, as we could detect qLMXBs with lower luminosities
in each of the other clusters.  We perform our qLMXB analysis both with and
without the (lower S/N) Terzan 5 data, using the faintest qLMXB
detected in Terzan 5 as its limiting luminosity.  The results of this
analysis are listed in Table 3, and the cumulative qLMXB luminosity
function for all clusters except Terzan 5 is shown in Figure 2.

For the harder, primarily CV sources above $10^{31}$ ergs s$^{-1}$, we use the
luminosities reported in GHE01a for 47 Tuc,  in
Pooley et al. (2002a) for NGC 6752, and in Heinke et al. (2003c) for
M80.    For NGC 6397, we have adjusted the luminosities calculated in
GHE01b for a distance of 2.7 kpc (Anthony-Twarog \& Twarog 2000). For
M28, we calculated the 0.5-2.5 
keV unabsorbed X-ray luminosities of 
sources within the half-mass radius of the cluster from the reported
 0.5-8.0 keV fits using PIMMS and the
spectral fits in Becker et al. (2003).  We use the MEKAL fits reported for M28
sources 17, 25, and 28, and the power-law fit for the remaining 11 sources
 (excluding the MSP). We have not removed the five known 
active binaries from this distribution, but we have removed the bright 
MSP in M28. We have not subtracted possible
background AGN, which should number no more than 1-2 among these
sources.    We show the cumulative LF for
the hard sources in these five clusters in Figure 2, along with the
LFs for the hard sources in 47 Tuc and NGC 6397 separately.  We also
analyze the combined  
LF above $5\times10^{31}$ ergs s$^{-1}$ for these five clusters plus NGC
6440 (Pooley et al. 2002a), which
suffers incompleteness and crowding at lower luminosities.  Finally,
we separately analyze the hard source LFs down to $10^{31}$ ergs s$^{-1}$ for
the five deep cluster observations (47 Tuc, NGC 6397, NGC 6752, M80, and M28).
The results of these analyses are 
listed in Table 3, while some of the KS probabilities as a function of
$\gamma$ are shown in Figure 3.

The index of the qLMXB luminosity function depends strongly upon the
 chosen minimum luminosity.  Assuming the qLMXBs are 10 km objects,
 the best-fit X-ray luminosity slope above the minimum is $\gamma=0.6$,
 while the bolometric luminosity slope is slightly steeper.
 The LFs of the hard source population in NGC
6397 and 47 Tuc are clearly distinct, and that of NGC 6397 differs from the
other globular clusters in general (Figures 2 and 3).  Since CVs make up most
 of these  
hard sources, there must be a difference in the CV properties between
these clusters.  (Removing the three known ABs among the 47 Tuc
 sources and the one known AB among the 6397 sources does not affect
 the results.)  We note that several cluster properties appear to
correlate with the slope of the hard source LFs; cluster central
density (inverse correlation), cluster metallicity, and cluster
collision frequency $\Gamma$ (see Table 1).  Since a large part of these
correlations are due to the unusual cluster NGC 6397 (see \S 4.3,
 4.4), we do not 
attempt to draw conclusions about which properties are responsible for
the different LF slopes.  

Possible high-luminosity cutoffs may exist in the luminosity functions
for both CVs and qLMXBs, as seen in Figure 2.  We judge these cutoffs
to be at roughly $L_X$(0.5-2.5 keV)=$2\times10^{32}$ ergs s$^{-1}$ for
the CVs, and $10^{33}$ ergs s$^{-1}$ for the qLMXBs.  KS tests show
both samples to be formally consistent with power-laws with no high-$L_X$
cutoffs. However, the KS probability decreases and 
$\gamma$ increases as the limiting luminosity increases, suggesting a
cutoff.   As the number of globular cluster X-ray sources identified in these
luminosity ranges increases, it will become worthwhile to fit more complicated
luminosity functions to the data.

\subsection{Dependence of X-ray Source Numbers upon Cluster Density and Core Radius}

We use the numbers of qLMXBs in several clusters of different structural
parameters (all the clusters from Table 1) to attempt to constrain qLMXB
and CV formation mechanisms.  We assume that dynamical formation of
globular cluster X-ray sources can be parametrized 
as $\Gamma=\rho_c^{\alpha}r_c^{\beta}$, where $\Gamma$ is the
formation rate, $\rho_c$ is the central luminosity density, and $r_c$
is the core radius.  We follow the density weighting method of Johnston \&
Verbunt (1996) to test the dependencies of qLMXB formation upon the
exponents $\alpha$ and $\beta$.  This method calculates the weight for
each cluster based upon the choice of $\alpha$ and $\beta$, and
assigns a line length proportional to that weight.  The line length is
reduced by the fraction of X-ray sources to our chosen limit that are
detectable in each cluster.  This only affects Terzan 5; using the
estimate of 30\% 
incompleteness above $10^{32}$ ergs s$^{-1}$ (due to the LMXB, Heinke et
al. 2003b), we find that 70\% of the qLMXBs in Terzan 5 should have
been detected, or 85\% if the transient is included.  The X-ray sources
within each cluster are spread evenly along the line segment, and then
the clusters are ordered by increasing weight to form a line of unit
length with all the
X-ray sources spread along it.  A KS test then is applied to check
whether the sources are consistent with a uniform distribution. 
The results of KS tests for a range of values of
$\alpha$ and $\beta$ are shown in Figure 4, with the best-fit
combination marked with a cross and 90\%, 50\%, and 10\% KS
probability contours marked.  The meaning of a KS probability P is
that a random distribution will be less uniform than the data P\% of
the time.  However, since we distribute the sources evenly within each
line segment, the KS probabilities will be overestimates.  Although
the results are not extremely 
constraining, clearly $\alpha$ is best fit near a value of 1.5. 

We perform the same analysis for harder sources above $10^{31}$
ergs s$^{-1}$ in a few clusters, which we believe are largely CVs, and the
subset of those sources above $10^{32}$ ergs s$^{-1}$ in a larger range of
clusters.  To avoid dependence of our results upon the assumed LF, we
use data that are complete to our chosen limiting luminosity except
for Terzan 5, where the incompleteness to $10^{32}$ ergs s$^{-1}$ is 30\%.
For hard sources above $10^{32}$ ergs s$^{-1}$, we use all the
clusters in Table 1 except NGC 6366 (for which cluster membership
is in doubt for all sources).  For hard sources above $10^{31}$
ergs s$^{-1}$, we exclude Terzan 5, NGC 6440, M13, and
NGC 6366 due to their incompleteness. For $\omega$ Cen the results of
Cool et al. (2002) and 
Gendre et al. (2003a) are in close agreement on the numbers of X-ray
sources above $10^{31}$ ergs s$^{-1}$ within the half-mass radius (15
vs. 16; source \# 9 brightened in the XMM data).  This sample 
includes only one soft source, the qLMXB, so 
foreground stars do not seem to be a problem. We estimate 4.7
background AGN from the analysis of Giacconi et al. (2001), and so
estimate 9 hard sources above $10^{31}$ ergs s$^{-1}$ belong to the cluster
(at the time of the \Chandra\ observation).
We also subtract one expected background source from NGC 6266 and
M28, and two expected background sources from 47 Tuc and NGC 5904,
producing the numbers listed under Hard Srcs in Table 1.  

 The results of these tests are shown in Figures 5 and 6, for the
bright and full sample of hard sources, respectively.  The
hard sources require lower values of $\alpha$ 
than the qLMXBs, and the sources above $10^{31}$ ergs s$^{-1}$ seem to
require values of $\beta$ less than 2.  This reduced dependence on the
core radius may be traced to the large number of CVs in NGC
6397, which is a core-collapsed cluster with a very small core
radius. We repeat the test for the full sample of hard sources
excluding NGC 6397, and find 
$\beta$ to be much more loosely constrained (Figure 7).  We discuss
these results in \S 4.3. 

\section{Discussion}

\subsection{Spatial distribution, spectra and variability}

The spatial distribution of our qLMXB sample agrees with our
expectation that these objects have masses characteristic of NSs plus
low-mass companions.  
The implied average qLMXB system mass of 1.5$^{+0.3}_{-0.2}$ (90\%
conf. errors) \Msun indicates 
that most NSs in qLMXBs have not accreted $\sim0.5$ \Msun from their
companions. This is in general agreement with estimates of the masses
of persistent LMXBs 
in several globular clusters (1.8$^{+1.2}_{-0.4}$ \Msun, Grindlay et al. 1984
recalculated in \S 3.1)
and possible MSPs in 47 Tuc, thought to be their descendants
(1.26$^{+0.14}_{-0.19}$ \Msun, Grindlay et al. 2002 recalculated in \S 3.1),
both derived using the same method.  It is also in agreement with the
masses of pulsars (and MSPs in particular) derived by Thorsett \&
Chakrabarty (1999), 1.35$\pm0.04$ \Msun, which indicates that very
little mass is required to spin neutron stars up to millisecond
periods.  Although our finding does not rule out the possibility that some 
qLMXBs may have larger masses, it lends weight against the high mass
interpretation of the bright qLMXB X7 in 47 Tuc (Heinke et al. 2003a).
As massive neutron stars should cool faster than less massive ones
(e.g. Colpi et al. 2001), the brightest qLMXB (X7) should not be the
most massive one.  

Most of the qLMXBs in Table 2 are consistent with a $\sim$10 km neutron
  star radius when fit with a hydrogen atmosphere model.  A smaller
  radius might suggest that a polar cap was strongly heated, perhaps
  through ongoing accretion channeled by a magnetic field.  A larger
  radius, as suggested for 47 Tuc X7 and M30 A1, can be explained by
  a more massive ($>1.7$\Msun) neutron star or by an alteration of
  opacity through 
  continued accretion (see Heinke et al 2003a, Lugger et al. 2003).
 Either possibility is of great interest.  We note that our method in
  this paper of
  calculating the errors on the radius of these qLMXBs underestimates
  the true errors (see in't Zand et al. 2001), and thus these radii 
  should be taken only as a measure of consistency with expectations,
  and not as rigorous constraints.  See Heinke et
al. (2003a, \& in prep.) and Lugger et al. (2003) for detailed
constraints on the radius and/or mass of some of these qLMXBs.   

The qLMXBs listed in Table 2 have very little or no power-law
component, in contrast to field systems identified through their
high-luminosity outbursts which require 10-40\% of their 0.5-10 keV
emission in this component.   Some
field qLMXBs that have recently been accreting (KS 1731-260, Wijnands et
al. 2002b and refs therein; MXB 1659-298, Wijnands et al. 2003) allow
a power-law component to constitute up to $\sim$25\% of the 0.5-10 keV
flux, but do not require it.  The only globular
cluster qLMXB to require this component, NGC
6440 CX1, is the only qLMXB among our sample to have experienced a
recorded outburst. 
This gives additional support to the suggestion by
Heinke et al. (2003a) that the strength of
the power-law component may be a measure of continuing low-level
accretion. We note that field qLMXBs that have shown intrinsic 
variability on short time scales (Cen X-4, Campana et al. 1997 and
Rutledge et al. 2003; Aql X-1,
Rutledge et al. 2002b) indicative of continued accretion clearly show this
power-law component.  However, M28 \#26 shows variability on a
timescale of months without evidence of this power-law component
(Becker et al. 2003), which does not fit this paradigm. However, this
variability may be due to changing $N_H$ column depth.  No
qLMXB without a power-law spectral component has yet been shown to require
variability in its thermal component, suggesting that the thermal
emission in these systems is entirely due to deep crustal heating
(Brown et al. 1998).

\subsection{Luminosity information}

We use the theoretical scaling of qLMXB formation with central density
and core radius (see \S 3.3, 4.3) to calculate relative $\Gamma$s for
the clusters 
in Table 1, as percentages of the total formation rate in the
galactic globular cluster system.  We can extrapolate from the studied
clusters to the remaining globular clusters in the catalog of Harris
(1996), and thus estimate  
that roughly 95 accreting neutron star systems may be found among the
entire galactic globular cluster system, in agreement with the 
estimate of Pooley et al. (2003b).  As discussed in \S
4.5 below, unusual dynamical histories of some clusters may
increase this number slightly.   Thirty-eight accreting
neutron star systems have now been identified (including the other
eleven bright LMXBs and the qLMXB in NGC 6652). Seven times more
qLMXBs in globular clusters are inferred than have been seen in outburst.  

Wijnands et al. (2001, 2002a, 2003) suggest that there exist a population of
 qLMXBs with long--duration outbursts ($>10$ years) and extremely long dormant
 periods (thousands of years).  This is required to explain the low
 quiescent flux from 
 several qLMXBs that have been accreting for many years, and thus
 would otherwise be expected to have very hot cores (Brown et
 al. 1998).  Pfahl, Rappaport, \& 
Podsiadlowski (2003) suggest that irradiation-induced mass transfer
cycles, with long periods of dormancy, may be required to account
 for the apparent lack of LMXBs compared to the number of MSPs in the galaxy. 
Our evidence that most accreting neutron stars in globular clusters
 are in deep quiescence supports the picture of long dormant
 periods suggested by Wijnands, Pfahl and others, although globular
 cluster systems may be very different from field systems.  

\Chandra\ observations have been sufficiently sensitive to observe soft
qLMXBs in 
many clusters below $1\times10^{32}$ ergs s$^{-1}$ (all clusters in Table
1 except Terzan 5, M13 and probably NGC 6440), but they have not
been seen.
Assuming 10 km radii in the spectral fits leads us to conclude that no
bolometric (redshifted) luminosities are below $2.3\times10^{32}$ ergs
s$^{-1}$.  
This lower LF cutoff implies a lower limit to the time-averaged mass
transfer rate of these systems in the Brown et al. (1998) model.
 If enhanced neutrino cooling is not active and the mass transfer is
mostly conservative, the bolometric
luminosity range of qLMXBs given in Table 2 translates to
time-averaged mass transfer rates of $3\times10^{-12}$ to
$7\times10^{-11}$ \Msun year$^{-1}$ (the latter for X7 in 47 Tuc).  

This range may represent the actual mass transfer rates, or may be an
underestimate due to enhanced neutrino cooling  
(e.g. Colpi et al. 2001) or nonconservative mass transfer.  We
note that the larger number is a factor of a few less than the disc stability
criterion for systems of periods similar to X5 (8 hours) in 47 Tuc
(King 2003).  Significantly larger mass transfer rates would cause
persistent emission instead of transient behavior.  
Taking these rates at face value, two explanations are plausible.  
Part of this range of mass transfer rates might be supplied by
initially evolved 
secondaries, i.e. originally intermediate-mass X-ray binaries,  as
suggested to predominate by recent work (Pfahl et al. 2003).
Alternatively, extremely old post-minimum systems will naturally
generate low mass-transfer rates below $10^{-11}$ \Msun year$^{-1}$
(King 2000).   In this picture, the lower luminosity limit is
attributed to the finite age of the systems, less than a Hubble time
(L. Bildsten, 2003, priv. comm.).  We note that this luminosity range
can be taken as support for the Brown et al. (1998) deep crustal
heating model.
An even lower thermal luminosity has been observed from the (low
mass-transfer, \mdot$\sim5\times10^{-12}$ \Msun year$^{-1}$) millisecond
X-ray pulsar SAX J1808, and has been taken to imply enhanced neutrino
cooling from the core (Campana et al. 2002).  If such enhanced cooling
is common among globular cluster qLMXB systems, the time-averaged mass
transfer rates may be higher than we have calculated.  

\subsection{Distribution among globular clusters}

Several methods of production of globular cluster binary X-ray sources
have been suggested. Normal evolution from primordial binaries is not a
reasonable explanation for neutron star systems (Clark 1975), but may
be able to explain some CVs and ABs (Verbunt \& Meylan 1988, Davies
1997). The numbers of such systems would depend upon the initial mass
of the cluster and the primordial binary fraction, and would be
suppressed by the destruction of wide binaries in close encounters
(Davies 1997).  Tidal capture of 
main-sequence stars by neutron stars or white dwarfs may generate
large numbers of short-period systems  (e.g. di
Stefano \& Rappaport 1994), while exchange encounters will tend to
inject neutron stars into longer-period primordial binaries (e.g. Hut,
Murphy \& Verbunt 1991). 
Both of the latter mechanisms predict a rate of formation of neutron star
binaries $\Gamma$ proportional to $\rho_c^2 r_c^3 /\sigma \propto
\rho_c^{1.5} r_c^2$, where $\rho_c$ is the central luminosity density,
$r_c$ is the core radius, and $\sigma$ the central velocity dispersion
(Verbunt \& Hut 1987, Verbunt 2003).    This 
assumes that the mass-to-light ratio is similar in the cores of
different globular clusters, and that most of the encounters happen
within the core.  The second assumption  
is a good approximation for King-model clusters due to the steep
density decline outside the core, but is less accurate for
core-collapsed clusters with a less steep decline outside the core.  A
more accurate 
calculation, as performed in Pooley et al. (2003b), 
integrates the density distribution out to large radii, using
the best surface profiles available.  Our method does allow us, however, to
investigate the dependence of X-ray source formation on core density
and core size separately. 

Our simple method does not account
for increased neutron star density in the core due to mass
segregation (Verbunt \& Meylan 1988), or for escape of neutron stars
from globular cluster potential wells of different depth (see Pfahl et
al. 2002).  It also carries the potential for significant bias in that
we only analyze a small sample of clusters, generally selected because of the
existence of known X-ray sources and perhaps not representative
of the general globular cluster system.  With these caveats, we
proceed. 

Our test of the distribution of globular cluster neutron star systems
in Figure 3 indeed shows compatibility with
$\Gamma\propto\rho_c^{1.5} r_c^2$, as predicted by either tidal
capture or exchange encounters.  This result agrees with the results
of Verbunt \& Hut (1987) on bright cluster LMXBs and the simpler tests
of Gendre et al. (2003b) and Pooley et al. (2003a, b) upon qLMXBs in
globular clusters.  We note that this result is also compatible with the
results of Johnston, Kulkarni \& Phinney (1992) on recycled pulsars
(the products of 
accreting neutron star systems) and those of Johnston \&
Verbunt (1996) on globular cluster low-luminosity X-ray sources (the
brightest of which are predominantly qLMXBs) when their neglect of the
velocity dispersion $\sigma\propto\rho^{0.5} r_c$ is considered
(as noted in Verbunt 2003). Grindlay (1996), using measured velocity
dispersions, 
indeed showed correlation of dim source numbers with the theoretical scaling.

For the harder sources (predominantly CVs), we do not find agreement
with the theoretically predicted 
formation rate from close encounters.  Figures 4, 5, and 6 show
a weaker dependence upon the central density for hard sources than for
qLMXBs.  The density dependence of the hard sources is better fit by
$\Gamma\propto\rho_c^{\alpha} r_c^2$ with $\alpha\sim$1.1--1.3, rather than
$\alpha\sim$1.5 as for qLMXBs.    One
suggestion for the lower density dependence is that high 
density environments destroy CVs preferentially, possibly by encounters
with neutron stars (e.g. Pooley et
al. 2002b, Verbunt 2003), especially during the core collapse
process.  NGC 6397, the densest globular cluster yet studied, has an
apparent excess of X-ray sources (Table 1, and Pooley et al. 2003b),
which suggests the opposite conclusion.  However, NGC 6397 may be
unusual for other reasons; see \S 4.4.

The formation of some CVs in globular clusters from primordial,
undisturbed binaries 
(especially in massive, moderately dense clusters like $\omega$ Cen;
Davies 1997, Verbunt 2003) may partially explain the weaker dependence
upon density.   
Many sources outside the half-mass radius of $\omega$ Cen may be 
CVs, two of them brighter than $L_X=10^{32}$ ergs s$^{-1}$ (Cool et al. 2002,
Gendre et al. 2003).  Davies (1997) indeed
predicts that CVs from primordial binaries should exist in the outer
regions of this unrelaxed globular cluster.  However, even if four
bright primordial CVs exist in $\omega$ Cen, the masses of the other globular
clusters in Table 1 are so much smaller that most of their bright CVs must
not be primordial.   According to Pryor \& Meylan (1993), $\omega$ Cen is as
massive as all the other globular clusters with bright CVs (except Ter
5 and NGC 6440) put together.  These other clusters have 21 bright
CVs, which shows that the majority of these CVs are not primordial.

\subsection{Other clusters and processes}
  
Our analysis of X-ray source distributions is consistent with X-ray
imaging studies of 
some dense high-$\Gamma$ clusters (Liller 1, Homer et al. 2001; M15,
Hannikainen et al. 2003) which suggest numerous X-ray
sources in the luminosity range of qLMXBs.  However, several  
clusters with low predicted collision rates seem to have more X-ray
sources than expected based on their structural parameters; besides
NGC 6397, these include NGC 6712, NGC 288, NGC 6652, and Terzan 1.  
Other dynamical processes may be at work in these clusters.

  Analysis of
the core-collapse process suggests that binaries in the core release
energy to passing stars by ``hardening'' into tighter orbits, thus
slowing the collapse process before their ultimate ejection or merger
(Hut et al. 1992, Fregeau 
et al. 2003).  The hardening of binaries
during core collapse might be expected to generate increased mass
transfer rates and thus X-ray emission.  As core collapse proceeds,
main-sequence binaries will be destroyed, but neutron stars should be
preferentially exchanged into binaries.
The numbers of each kind of X-ray binary existing at any one time 
may thus be a function of the cluster dynamical history, as well as 
current structure.  Differences in dynamical history may  
explain the unusually large numbers of CVs (and NS systems?) in NGC
6397 (see Table 
1), compared to the similar core-collapsed globular clusters M30 and
NGC 6544.  The 
latter, while not yet surveyed by \Chandra, has a ROSAT upper limit of
$L_X=5.9\times10^{31}$ ergs s$^{-1}$ (Verbunt 2001) above which 5 sources
exist in NGC 6397.  NGC 6397 has recently been shown to be depleted in
binaries compared 
to several other clusters (Cool \& Bolton 2002), while it appears to
have an abundance of X-ray sources.  NGC 6397 also seems to show an
unusually flat CV LF, compared to 47 Tuc and other clusters in general
(\S 3.2, and Pooley et al. 2002b).

The cluster NGC 6712 (which contains a bright persistent ultracompact
LMXB) shows evidence 
for multiple accreting binaries and blue stragglers despite its
relatively low density and 
$\Gamma$ (Ferraro et al.\ 2000, Paltrinieri et al. 2001;
$\Gamma=0.13$).  The path of NGC 6712's orbit through the galactic 
bulge suggests that it experiences strong tidal stripping from the
galactic potential (Dauphole et al. 1996). NGC 6712's  
declining mass function below the turnoff (de Marchi et al. 1999,
Andreuzzi et al. 2001) gives strong
evidence that it has been 
stripped of $\geq99$\% of its initial mass (Takahashi \& Portegies
Zwart 2000).  An analysis of the orbits and structure of 38 globular
clusters by Dinescu, Girard \& van Altena (1999) indicates that
NGC 6712, NGC 6397, NGC 6121, NGC 288, Palomar 5, and possibly M80
have very high destruction rates.  
Ferraro et al. (2000) thus suggest that NGC 6712 is only
the fossil remnant core of a once very massive cluster, heavily
enriched in compact objects and binaries due to mass segregation.
Disrupting globular clusters  should be 
generally marked by an apparent excess of massive stars, binaries, and
binary products (such as X-ray sources or blue stragglers), which will
remain segregated in the core while the outer halo is stripped.
This destruction 
 process should deposit X-ray binaries from globular clusters into the
 galactic bulge (Grindlay 1985).

NGC 6652 and Terzan 1 each possess a bright LMXB, although their
 collision rates are very low ($\Gamma=0.18$ and 0.008 respectively).  In
 addition, they are home to at least three and one additional X-ray
 sources above $\sim5\times10^{32}$ ergs s$^{-1}$, respectively (Heinke et
 al. 2001; Wijnands et al. 2002a).   The orbits  
 of NGC 6652 and Terzan 1 have not yet been calculated, but Idiart et
 al. (2002) note that Terzan 1 seems to have captured metal-rich stars
 from the bulge, implying it was once much more massive.  Two of NGC 6652's
 X-ray sources, both probable neutron star systems, are well outside
 the core (Heinke et al. 2001). Although NGC 6652 does not
 show signs of core collapse, this implies an unusual dynamical state.
  Analysis of archival
 \HST\ images to measure the stellar LF and surface profile of NGC 6652
 could test this.  NGC 6652 and Terzan 1 could be remnant cores of
 initially much more massive 
 globular clusters on their way to destruction.

Several of the other high-destruction clusters show low-mass star
  depletion (NGC 6121, Kanatas et al. 1995; NGC 288, Bellazzini et al.
  2002b; NGC 6397, Piotto, Cool \& King 1997).  Palomar 5 shows both
  low-mass star depletion (Grillmair \& Smith 2001) and tidal tails
  (Odenkirchen et al. 2001).  These facts suggest that cluster destruction 
 processes may have concentrated these clusters' X-ray populations as
  well, although not as severely as NGC 6712.  Such processes have
  recently been 
  suggested by Pooley et al. (2003b) to explain the apparent excess of
  X-ray sources in NGC 6397.  Disentangling the effects of
  cluster destruction and core collapse on binary production will
  probably require significant theoretical work.  

The very loose cluster NGC 288 has an X-ray source with 
$L_X\sim3\times10^{32}$ from ROSAT HRI data (Verbunt 2001), even though
it is very poor with $\Gamma$=0.005.  NGC 288 shows centrally
concentrated binary stars and blue stragglers (Bellazzini et
al. 2002a; Ferraro et al. 2003).  
However, Bellazzini et al. (2002a) think that it would be very
 difficult to re-expand NGC 288 to its current low density after
 compressing the core sufficiently to produce collisional products.  They and
 Ferraro et al. (2003) instead suggest that 
 NGC 288's large numbers of blue stragglers are the result of
 primordial binary evolution, like the halo blue stragglers in M3
 (Ferraro et al. 1997), and may be due to an initially high binary
 fraction.  The high destruction rate of NGC 288 suggests that its 
 binary fraction may be enhanced by the removal of lower-mass single
 stars from the halo during its repeated disk shocks.  It will be of
 great interest to see if the same processes that generate large
 numbers of blue stragglers in NGC 288 also produce numerous X-ray
sources detectable in recent \Chandra\ observations (principal
investigator: W.~G.~H. Lewin).   

Metallicity has been suggested to have an effect on the luminosity of
extragalactic globular cluster X-ray sources (e.g. Kundu, Maccarone, \& Zepf
2002). This metallicity effect may be due to differences in cluster
initial mass functions (Grindlay 1993) or to the larger radii of
metal-rich stars 
increasing both tidal capture and Roche-lobe overflow rates
(Bellazzini et al. 1995).   More massive clusters should be better able
to retain neutron stars after their formation kicks (Pfahl et
al. 2002), which will affect the numbers of MSPs as well as qLMXBs.
Neither mass nor metallicity have obvious effects upon the galactic
globular clusters studied here.   
Analyzing the effects of all the above factors on the various globular
cluster populations will require deep observations of a number of
clusters with very different parameters, paired with deep radio and
optical datasets to clearly identify different source types below the
luminosities discussed here.

\section{Conclusions}

We have created a catalog of known qLMXBs containing neutron stars in
globular clusters, adding three probable qLMXBs in NGC 6440 to those
already known.  We have reanalyzed those qLMXBs in archived \Chandra
globular cluster observations using the hydrogen atmosphere models of Lloyd 
 (2003), and find general consistency with 10 km radii.  The hard
power-law component required in the spectra of many field qLMXBs is
absent in most globular cluster qLMXBs, with the notable exception of
the recently active transient in NGC 6440.  The radial distribution of
these qLMXBs within their globular clusters is consistent with a mass
of 1.5$^{+0.3}_{-0.2}$ \Msun.  This is as expected for accreting neutron
star systems, and suggests that the neutron stars do not grow
significantly in mass.   Globular cluster qLMXBs
range in luminosity from $10^{32}$ ergs s$^{-1}$ up to a few $10^{33}$
ergs s$^{-1}$.  Quiescent LMXBs below $10^{32}$ ergs s$^{-1}$ would have been
identifiable in most clusters, so the cutoff implies a lower limit to
the time-averaged mass accretion rate.  This range of luminosities is
consistent with the Brown et al. (1998) model for qLMXB emission, as
higher mass transfer rates would lead to persistent systems and
significantly lower
rates would probably require systems older than a Hubble time.

The luminosity function of globular cluster qLMXBs is consistent with
the LF of globular cluster sources analyzed by Johnston \& Verbunt
(1996), which is not surprising as they are usually the brightest
sources in a cluster.  The LFs of harder sources above $10^{31}$
ergs s$^{-1}$ , which are mostly CVs, appear to vary between clusters,
suggesting an influence of metallicity or core collapse upon CV
properties.  The numbers of qLMXBs in
different clusters are consistent with the relative numbers of close
encounters, allowing either tidal capture or exchange encounters as a
mode of production. This suggests that the total number of accreting
neutron stars in globular clusters is near 100.  The harder
sources, however, show a lesser dependence upon density,
suggesting that dense environments may tend to destroy CVs. The core
collapse process could be responsible for differences in the numbers
and types of X-ray
binaries between NGC 6397 and similar clusters. 
Tidal destruction or evaporation of clusters may leave substantial
numbers of X-ray sources in apparently poor clusters.  

This study has begun to probe individual source populations across
different globular clusters, and test the effects of varying central
density and core radius upon the properties of two populations, qLMXBs
and CVs.  Future \Chandra\ observations will allow us to test the
effects of metallicity, cluster mass, and other dynamical processes in
clusters.  Deep optical and radio datasets are also allowing
identification and study of populations of ABs and MSPs.  In addition
to understanding the properties of these binaries, this work offers an
opportunity for a deeper understanding of globular cluster evolution.

\acknowledgments

We are very grateful to D. Pooley for communicating results of several globular
cluster studies to us before publication.  We also thank L. Bildsten
for useful discussions, and A. Kong and D. Pooley for comments on the
manuscript.  C.~O.~H. acknowledges support from \Chandra\ grant GO2-3059A.


\begin{deluxetable}{lccccccccr}
\tablewidth{7truein}
\tablecaption{\textbf{Recent \Chandra\ or XMM Globular Cluster Surveys}}
\tablehead{
\colhead{\textbf{Cluster}} & \colhead{N$_H$} & \colhead{D} &
\colhead{log($\rho_0$)\tablenotemark{a}} & \colhead{r$_c$} &
  \colhead{$\Gamma$} &
\colhead{\# NS} & \multicolumn{2}{c}{\# Hard Srcs\tablenotemark{b}} & \colhead{Ref.} \\
 & (10$^{22}$ cm$^{-2}$) & (kpc) & ($\Lsun$ pc$^3$) & (arcmin) & (\% Gal.) &
 & ($>10^{32}$) & ($>10^{31}$) & 
}
\startdata
NGC~6440  	& 0.59 & 8.5 & 5.28 & 0.13 	& 6.4 & 8 & 5 & $>16$ & 1, 2  \\
Ter 5   	& 1.20 & 8.7 & 5.23 & 0.13 	& 5.9 & $>$5 & $>$4 & ? & 3, 4  \\
NGC~6266  	& 0.26 & 6.9 & 5.14c? & 0.18 	& 5.2 & 5 & 5 & 26$^b$ & 5, 6  \\
47 Tuc   	& 0.030 & 4.5 & 4.82 & 0.40 	& 3.6 & 2 & 3 & 22$^b$ & 7, 8  \\
M80   	       	& 0.094 & 10.3 & 4.87 & 0.11 	& 1.7 & 2 & 3 & 14 & 9  \\
M28   		& 0.24 & 5.5 & 4.73 & 0.24 	& 1.5 & 1 & 2 & 14$^b$ & 10  \\
NGC~6752      & 0.022 & 4.1 & 4.91c? & 0.17 	& 0.74 & 0 & 1 & 8 & 11, 3  \\
$\omega$ Cen   	& 0.09 & 5.3 & 3.03 & 3.15 	& 0.64 & 1 & 2 & $\sim$9$^b$ & 12, 13, 14, 15 \\
M30   	     	& 0.017 & 9.8 & 5.04c & 0.06 	& 0.58 & 1 & 0 & 3 & 16  \\
NGC~5904  	& 0.017 & 7.5 & 3.91 & 0.42 	& 0.48 & 0 & 0 & 4$^b$ & 5  \\
NGC~6397  	& 0.10 & 2.7 & 5.41c & 0.08 	& 0.40 & 1 & 3 & 8 &
17, 18, 19 \\
M22   		& 0.22 & 3.2 & 3.64 & 1.42 	& 0.39 & 0 & 0 & 3 & 20  \\
M13   		& 0.01 & 7.7 & 3.33 & 0.78 	& 0.23 & 1 & 2 & ? & 21, 22 \\
NGC~6121  	& 0.20 & 2.2 & 3.82 & 0.83 	& 0.12 & 0 & 0 & 1 & 23, 5  \\
NGC~6366  	& 0.39 & 3.6 & 2.42 & 1.83 	& 0.01 & 0 & 0-1 & ? & 5, 21 \\

\enddata
\tablecomments{Distances, reddening, core radius and cluster central
density taken from the most recent X-ray analysis work, or from the
Harris (1996) catalog (updated Feb. 2003) with a few updates.
Central densities are recalculated from prescription of Djorgovski
(1993). Numbers of sources from the quoted X-ray 
analyses inside the cluster half-mass radii, with luminosities in the
0.5-2.5 keV range.  NS refers to all 
accreting neutron star systems, including transients in NGC 6440 and
Terzan 5.  Close encounter rate $\Gamma$, calculated by
$\Gamma\propto\rho_0^{1.5} r_c^2$, as percentage of total galactic
globular cluster system rate. 
References: (1) Pooley et al. 2002b, (2) this work, (3) Cohn et
al. 2002, (4) Heinke et al. 2003b, (5) Pooley et al. 2003a, (6) Pooley
et al. 2003c (in prep), 
(7) Grindlay et al. 2001a, (8) Heinke et al. in prep, (9) Heinke et
al. 2003c, (10) Becker et al. 2003, (11) Pooley et al. 2002a, (12) 
  Rutledge et al. 2002, (13) Cool et al. 2002, (14) Gendre et
al. 2003a, (15) van Leeuwen \& Le Poole 2002, (16) Lugger et al. 2003,
(17) Grindlay et al. 2001b,  (18) Anthony-Twarog \& Twarog 2000, 
(19) Sosin 1997, (20) Webb et al. 2002, 
(21) Gendre et al. 2003b, (22) Verbunt 2001, (23) Bassa et al. 2003.
}
\tablenotetext{a}{c=core collapsed.  NGC 6752 (Lugger et
al. 1995.) and NGC 6266 (Harris 1996) may not be core-collapsed.}
\tablenotetext{b}{Including subtraction of probable background
sources, based on Giacconi et al. (2001) log N-log S.  We have
subtracted 1 hard source from NGC 6266, 2 from 47 Tuc, 1 from M28, 5 from
$\omega$ Cen, and 2 from NGC 5904.}
\end{deluxetable}

\begin{rotate}
\begin{deluxetable}{lccccccccr}
\tablewidth{8.4truein}
\tablecaption{\textbf{Probable qLMXBs in Globular Clusters}}
\tablehead{
\colhead{\textbf{Cluster, ID}} & \colhead{Cts} &  \colhead{Radius} & \colhead{kT$_{\rm eff}$} &
\colhead{N$_H$} & \colhead{$\chi^2_{\nu}$/dof}  &
\colhead{PL flux \%} & \multicolumn{2}{c}{$L^{\infty}$, ergs  s$^{-1}$} & \colhead{Ref.} \\
 & & (km) & (eV) & (10$^{22}$ cm$^{-2}$) &  & (0.5-10 keV) &
(0.5-2.5 keV) & (0.01-10 keV)  &
}
\startdata

47 Tuc, X5 & 4181  & 12.0$^{+7.5}_{-3.5}$  & 119$^{+21}_{-18}$ &
0.09$^{+.05}_{-.05}$ & 1.38/26 & 0$^{+3}_{-0}$ & 1.5$\times10^{33}$ &
2.2$\times10^{33}$ & 1, 2 \\

47 Tuc, X7 & 5508 & 35$^{+22}_{-12}$ & 84$^{+13}_{-12}$ &
0.13$^{+.06}_{-.04}$ & 1.20/26 & 0$^{+0.5}_{-0}$ & 2.0$\times10^{33}$
& 3.6$\times10^{33}$ & 1, 2 \\

$\omega$ Cen, \#3 & 467 & 12.4$^{+3.9}_{-3.1}$ & 77$^{+8}_{-7}$ & (0.09) &
0.90/18 & 0$^{+12}_{-0}$  & $1.9\pm.1(1.8)\times10^{32}$ &
4.1(3.8)$\times10^{32}$ & 3, 12 \\

6397, U24 & 660 & 10.0$\pm1.6$ & 74$^{+7}_{-6}$ &
(0.10)  & 0.88/23 & 0$^{+5}_{-0}$ &
1.0$\pm.03\times10^{32}$ & 2.3$\times10^{32}$ & 4, 12 \\

6440, CX1 & 247 & 5.9$^{+4.2}_{-1.4}$ & 150$^{+31}_{-19}$ & (0.59) & 0.47/10 &
28$^{+13}_{-15}$ & $1.1\pm.1(1.3)\times10^{33}$ &  1.4(1.6)$\times10^{33}$  & 5, 6, 12 \\

6440, CX2 & 172 & 6.9$^{+6.4}_{-2.4}$  & 137$^{+26}_{-26}$ & (0.59) &
1.13/6 & 0$^{+19}_{0}$  & $9.0\pm.7(9.8)\times10^{32}$ &
1.3(1.5)$\times10^{33}$ & 6, 12 \\

6440, CX3 & 116 & 9.7$^{+12}_{-1.7}$ & 108$^{+39}_{-15}$ & (0.59) &
0.81/3 & 18$^{+16}_{-17}$ & 6.6$\pm.6\times10^{32}$  &
1.0$\times10^{33}$ & 6, 12 \\

6440, CX5 & 90 & 9.2$^{+0.2}_{-4.4}$ & 109$^{+30}_{-26}$ & (0.59) & 63\% &
5$^{+11}_{-5}$ & $5.7\pm.6(5.9)\times10^{32}$ & $9.0(9.4)\times10^{32}$ & 6, 12 \\

6440, CX7 & 43 & 3.6$^{+5.0}_{-2.0}$ & 138$^{+58}_{-38}$ &
(0.059) & 49\% & 0$^{+15}_{-0}$ & $2.1\pm.3(3.0)\times10^{32}$
& $4.5(5.5)\times10^{32}$ & 6, 12 \\

6440, CX10 & 17 & 49$^{+178}_{-20}$ & 48$^{+31}_{-17}$ & (0.59) & 42\% &
0$^{+10}_{-0}$ & $2.1\pm.5(1.3)\times10^{32}$ & $9.4(2.9)\times10^{32}$ & 6, 12 \\

6440, CX12 & 12 & 6.8$^{+48}_{-5.3}$ & 86$^{+69}_{-44}$ & (0.59) & 50\% &
0$^{+42}_{-0}$ & $9.0\pm2.4(11.1)\times10^{31}$ & $1.7(2.5)\times10^{32}$ & 6, 12 \\

6440, CX13 & 11 & 2.2$^{+20}_{-0.8}$ & 127$^{+99}_{-72}$ & (0.59) & 47\% &
0$^{+60}_{-0}$ & $5.7\pm1.6(10)\times10^{31}$ & 8.6(23)$\times10^{31}$ & 6, 12 \\

Ter 5, W2 & 37 & 3.1$^{+15}_{-2.3}$ & 154$^{+118}_{-73}$ & (1.2) & 0.63/5 &
0$^{+36}_{-0}$ & $3.2\pm.8(4.5)\times10^{32}$ & $4.3(7.8)\times10^{32}$ & 7, 12 \\

Ter 5, W3 & 77 & 1.6$^{+3.1}_{-0.6}$ & 222$^{+50*}_{-78}$ & (1.2) & 0.71/9 &
0$^{+53}_{-0}$ & $4.0\pm.6(6.5)\times10^{32}$ & 5.1(10.6)$\times10^{32}$ & 7, 12 \\

Ter 5, W4 & 17 & 2.9$^{+280}_{-1.1}$ & 143$^{+127*}_{-115}$  & (1.2) & 1.49/11 & - &
$2.1\pm.9(2.2)\times10^{32}$ & $3.0(4.2)\times10^{32}$ & 7, 12 \\

Ter 5, W8 & 34 & 1.6$^{+280}_{-1.1}$ & 189$^{+83*}_{-132}$ & (1.2) & 0.43/6 &
3$^{+120}_{-3}$ & $2.1\pm.7(3.6)\times10^{32}$ & 3.0(6.4)$\times10^{32}$ & 7, 12 \\

M28 \#24 & 1669 & 11.1$^{+5.3}_{-2.9}$ & 118$^{+39}_{-14}$ & 0.26$\pm0.04$ &
0.96/44 & - &  $1.2^{+.7}_{-.4}\times10^{33}$ & $1.9\times10^{33}$ & 8 \\

M13 Ga & - & 9.8$\pm0.3$ & 99$\pm4$ & 0.11 & 0.55/15 & - & $4.3\pm.4\times10^{32}$ & $7.4\times10^{32}$ & 9 \\

M30, A1 & 830 & 20.5$^{+22}_{-5.7}$ & 86$^{+11}_{-13}$ &
0.034$^{+.055}_{-.018*}$ & 1.06/33 & 0$^{+6}_{0}$ & $8.8\pm.3\times10^{32}$ & $1.8\times10^{33}$ & 10 \\

M80, CX2 & 227 & 9.3$^{+3.2}_{-1.5}$ & 92$^{+13}_{-10}$ & (0.094) & 1.12/8 &
0$^{+12}_{-0}$  & $2.8\pm.2(2.8)\times10^{32}$ & 5.1(5.2)$\times10^{32}$ & 11, 12 \\

M80, CX6 & 62 & 4.2$^{+1.9}_{-0.5}$ & 95$^{+22}_{-4}$ & (0.094) & 50\%
& 0$^{+22}_{-0}$  & $8.0\pm1.0(12.7)\times10^{31}$ & 1.4(2.8)$\times10^{32}$ & 11, 12 \\
\enddata
\tablecomments{Parameters derived from spectral fits to probable
qLMXBs in globular clusters.  Spectral fits in XSPEC, using models of
Lloyd (2003; except for M28 and M13, see text) with grav. redshift
fixed to 0.306 (except for 47 Tuc, M30; see refs).  Photoelectric
absorption fixed at the known galactic absorption to the
cluster for the fainter sources, and equal or greater than the
galactic absorption for brighter sources.  Parameter values for CX5, CX7, CX10,
CX12, CX13 in 6440 and CX6 in M80 
derived using C statistic in XSPEC. For C statistic fits, the
percentage of Monte-Carlo simulations generating a C statistic less
than the best fit are given in place of the $\chi^2$ statistic.
 Luminosities
are given from the best fit and, where the fit is reasonable, for a
fit with radius forced to 10 km (in parentheses). All errors are 90\%
confidence limits on one parameter, except for X-ray luminosity errors 
which are 1$\sigma$ errors derived from counting statistics, ignoring
uncertainty in other parameters. A * indicates the fit encountered a
hard limit.  For M80 CX6, $N_H$
must be freed to allow an acceptable fit with radius=10 km, giving
$N_{H,22}=0.21^{+.05}_{-.10}$. 
References: (1) Grindlay et al. 2001a, (2) Heinke et 
al. 2003a, (3) Rutledge et al. 2002a,  (4)
Grindlay et al. 2001b, (5) Pooley et al. 2001b, (6) in't Zand et
al. 2001, (7) Heinke et al. 2003b, (8) Becker et al. 2003,
(9) Gendre, Barret, \& Webb 2003, (10) Lugger et al. 2003 (in prep),
(11) Heinke et al. 2003c,  (12) re-analyzed in this work. 
}
\end{deluxetable}
\end{rotate}

\begin{deluxetable}{lcccr}
\tablewidth{3.5truein}
\tablecaption{\textbf{Luminosity Function Fits}}
\tablehead{
\colhead{\textbf{Description}} & \colhead{$L_{min}$}  &
\colhead{\# srcs} & 
 \colhead{$\gamma$} & \colhead{Prob} \\
}
\startdata
\multicolumn{5}{c}{qLMXBs} \\
\hline
$L_X$, R fixed & $1.0\times10^{32}$ & 17  & 0.64$^{+.31}_{-.22}$ & 81\% \\
... with Ter 5 & $1.0\times10^{32}$ & 21 & $0.76^{+.36}_{-.30}$ & 90\% \\
$L_X$, R free & $5.7\times10^{31}$ & 17 & $0.46^{+.15}_{-.10}$ & 49\% \\
$L_{Bol}$, R fixed & $2.3\times10^{32}$ & 17 & $0.77^{+.43}_{-.29}$ & 91\% \\
... with Ter 5 & $2.3\times10^{32}$ & 21 & $0.94^{+.48}_{-.41}$ & 93\% \\
\hline
\multicolumn{5}{c}{Harder Sources} \\
\hline
Only 47 Tuc &  $1\times10^{31}$  & 18 & $0.85^{+.61}_{-.37}$ & 99\% \\
Only M80 &  $1\times10^{31}$  & 14 & $0.65^{+.51}_{-.30}$ & 99\% \\
Only M28 &  $1\times10^{31}$  & 14 & $0.79^{+.64}_{-.24}$ & 66\% \\
Only N6752 &  $1\times10^{31}$  & 8 & $0.62^{+.57}_{-.27}$ & 85\% \\
Only N6397 &  $1\times10^{31}$  & 8 & $0.42^{+.29}_{-.13}$ & 54\% \\
5 clusters & $1\times10^{31}$  & 62 & $0.67^{+.20}_{-.12}$ & 80\% \\
6 clusters &  $5\times10^{31}$  & 25 & $0.82^{+.19}_{-.15}$ & 42\% \\
\enddata
\tablecomments{ Results of KS tests for the LFs of source populations
in several globular clusters.  For qLMXBs, sources include those in
Table 2 omitting the four Terzan 5 sources, except where indicated.
For harder sources, the 5 clusters are those listed individually, with
NGC 6440 added to study only the bright CVs (6 clusters). $L_{min}$ is
the limiting 
luminosity of each analysis (except for Terzan 5 qLMXBs, where
$L_{X,{\rm min}}=2\times10^{32}$ and $4\times10^{32}$ ergs s$^{-1}$ for X-ray
and bolometric luminosities).  $\gamma$
is the index of the best-fit LF ($dN/dL_X \propto L_X^{-(\gamma+1)}$),
and the listed errors are the range where the KS probability (Prob) is larger
than 10\% (where the KS probability is the probability that a random sample
selected from the given distribution will have a larger statistic).
}
\end{deluxetable}

\clearpage


\psfig{file=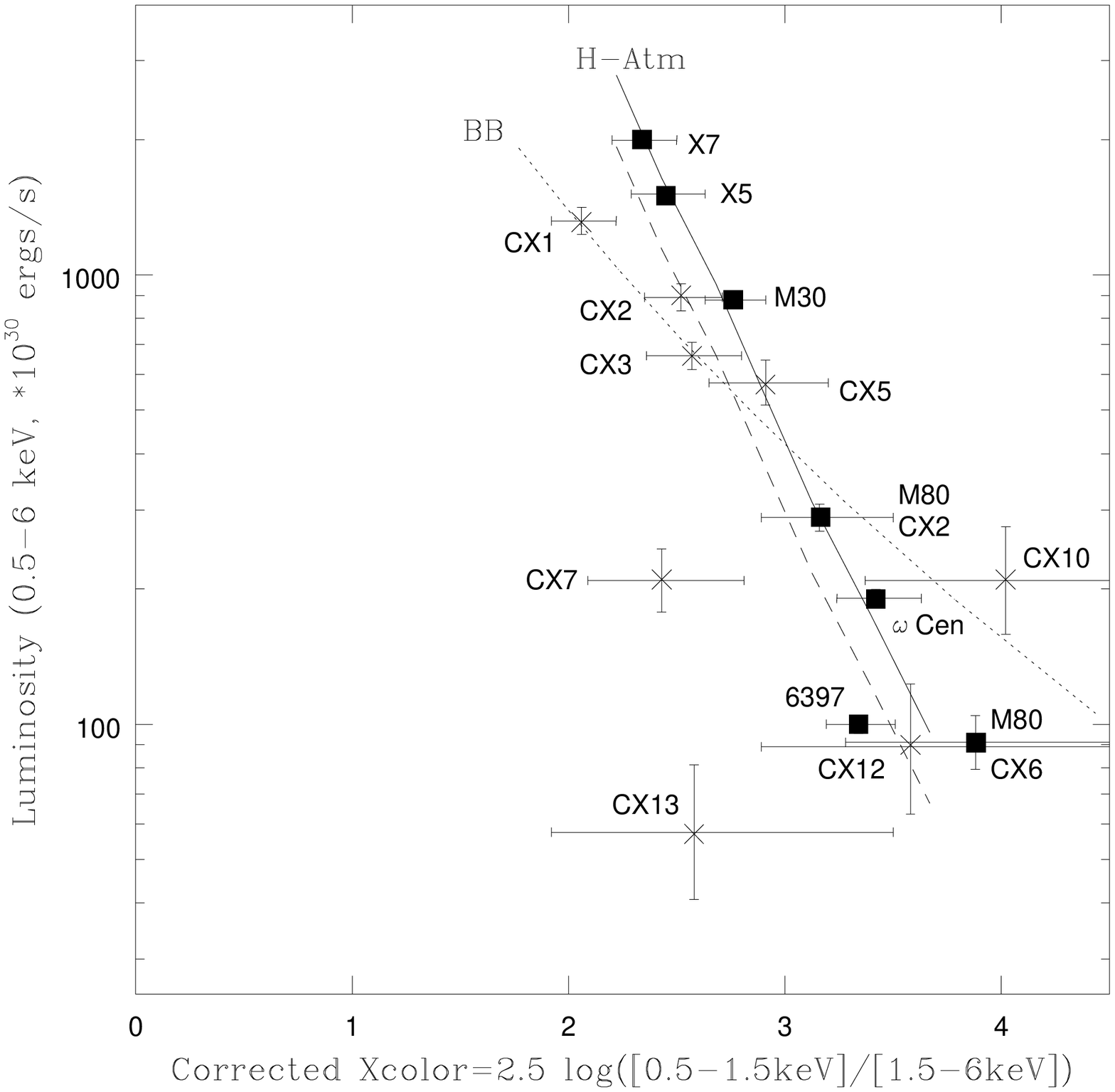,height=6in}

\figcaption[f1.eps]{Standardized X-ray CMD with several qLMXBs
and theoretical cooling tracks plotted.  Crosses represent NGC 6440
probable qLMXBs, while squares represent qLMXBs from 
other clusters (NGC 6397, $\omega$ Cen, M30, M80 CX2 and CX6, and  47
Tuc X5, X7). 
A theoretical 1.44 km blackbody track is plotted (dotted line), as are
10 (dashed) and 12 km (solid) nonmagnetic hydrogen-atmosphere models
from Lloyd (2003).  Errors are from Gehrels (1986) applied to the
counts detected in each band, and do not include possible errors in
the best-fit luminosity from spectral fitting.  CVs (not shown) tend to have
corrected Xcolors near zero on this scale. 
}

\psfig{file=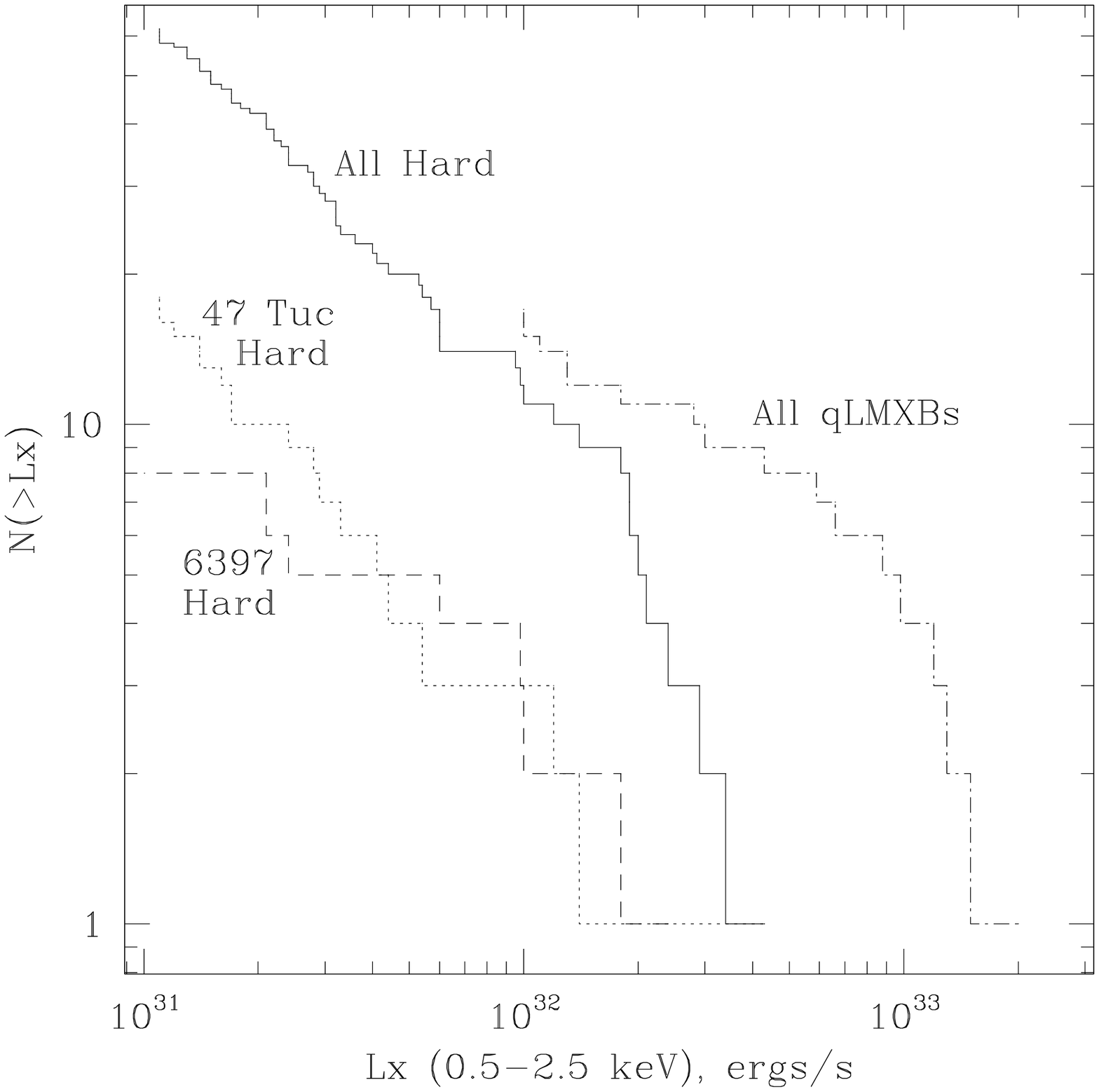,height=6in}

\figcaption[f2.eps]{Cumulative luminosity functions of qLMXBs
and probable CVs in globular clusters.  The qLMXBs are from Table 2,
using $L_X$s derived from fits assuming 10 km radius (where
acceptable). 
The probable CVs are hard sources with $10^{31}<L_X(0.5-2.5)<10^{33}$
ergs s$^{-1}$, with luminosities from 
47 Tuc (GHE01a), NGC 6397 (GHE01b), NGC 6752 (Pooley 
et al. 2002a), M28 (Becker et al. 2002), and M80 (Heinke et
al. 2003c). A few known active binaries have not been removed from the
CV sample, but the bright MSP in M28 has been removed.  One or two
background AGN are also expected among these sources.
}

\psfig{file=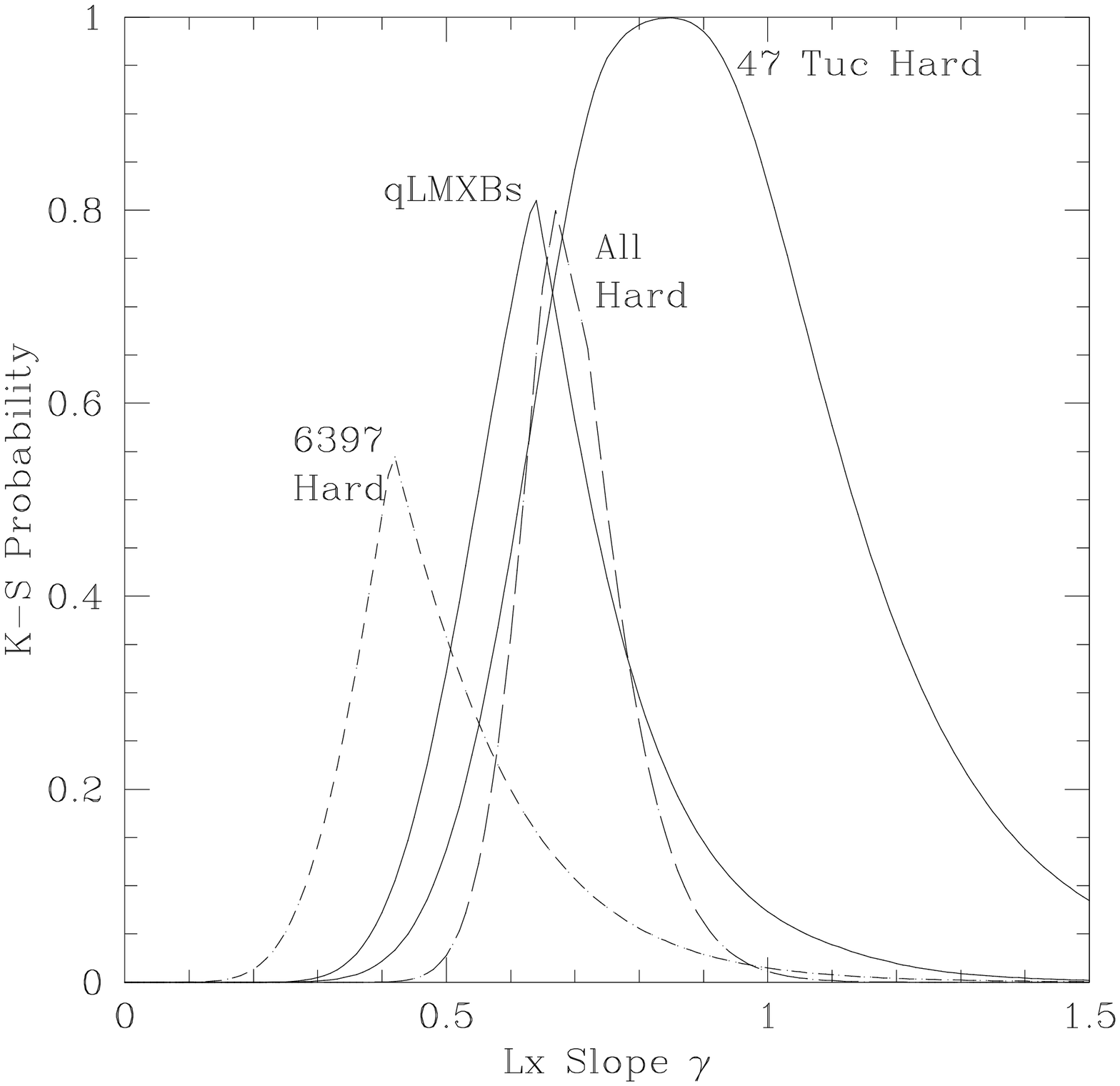,height=6in}

\figcaption[f3.eps]{Comparison of the indices of power-law
slopes for several different globular cluster populations.
Kolmogorov-Smirnov probability is plotted as a function of $\gamma$,
where the LF is assumed to be $dN \propto L_X^{-\gamma} d {\rm ln}
L_X$. The qLMXB LF is that where qLMXB radii are assumed
to be 10 km, and Terzan 5 qLMXBs are excluded (line 1 in Table 3).
}

\psfig{file=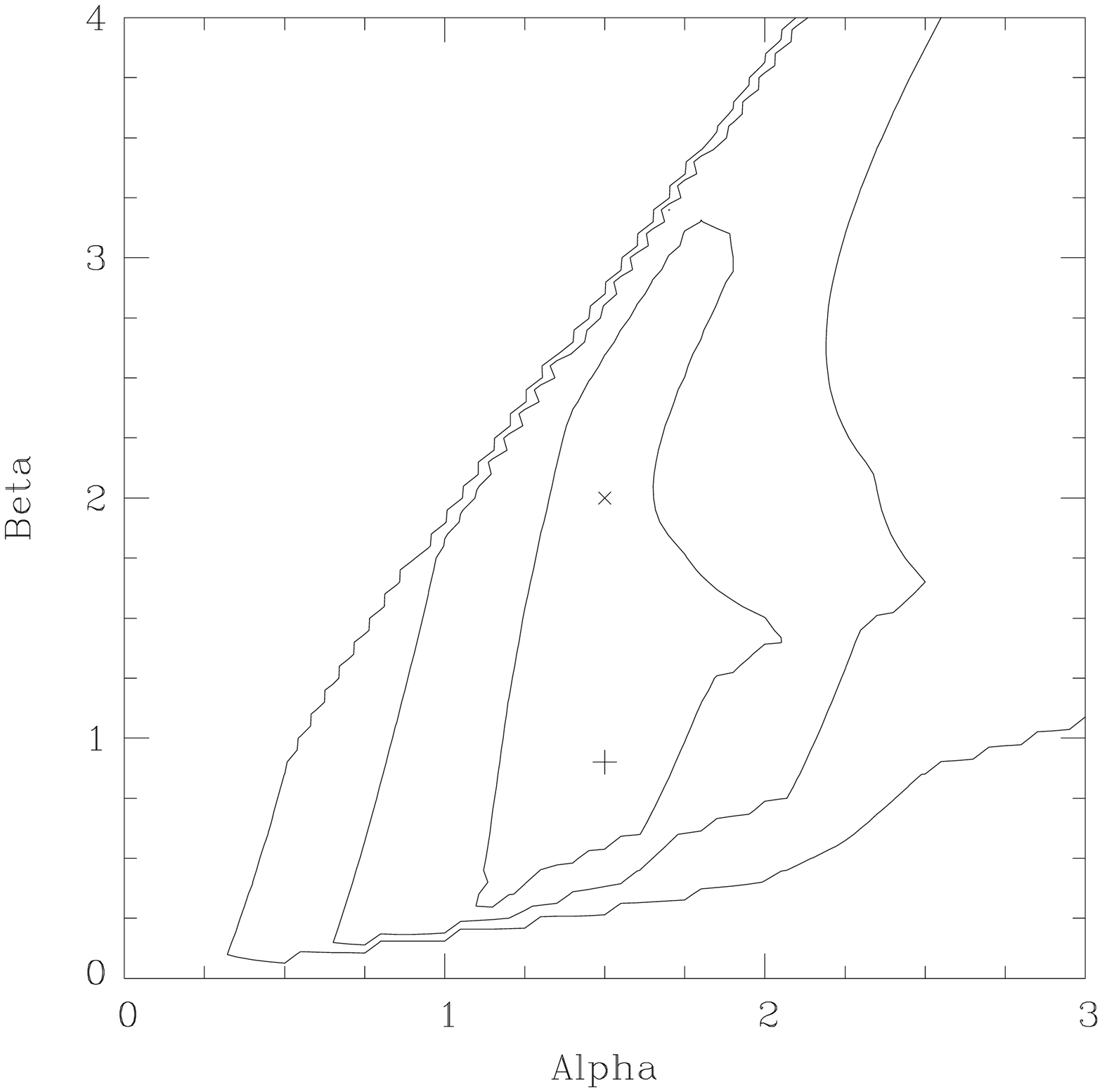,height=6in}

\figcaption[f4.eps]{Kolmogorov-Smirnov probability contour
map for the dependencies of qLMXB distribution among globular 
clusters upon central cluster density $\rho_0$ and core radius $r_c$.
The data are the numbers of
accreting neutron stars (whether in outburst or quiescence) from all
clusters in Table 1, plus an estimate of the 
incompleteness of the Terzan 5 survey (see text). We test the
acceptability of a qLMXB formation rate 
$\Gamma$ of the form $\Gamma \propto \rho_0^{\alpha} r_c^{\beta}$ for
various values of $\alpha$ and $\beta$.   Contours indicate K-S
probabilities of an acceptable distribution of 10\%, 50\%, and 90\%,
while the cross (+) marks the best fit.  The theoretically calculated
close encounter rate dependency (Verbunt \& Hut 1987) is indicated by an X. 
}

\psfig{file=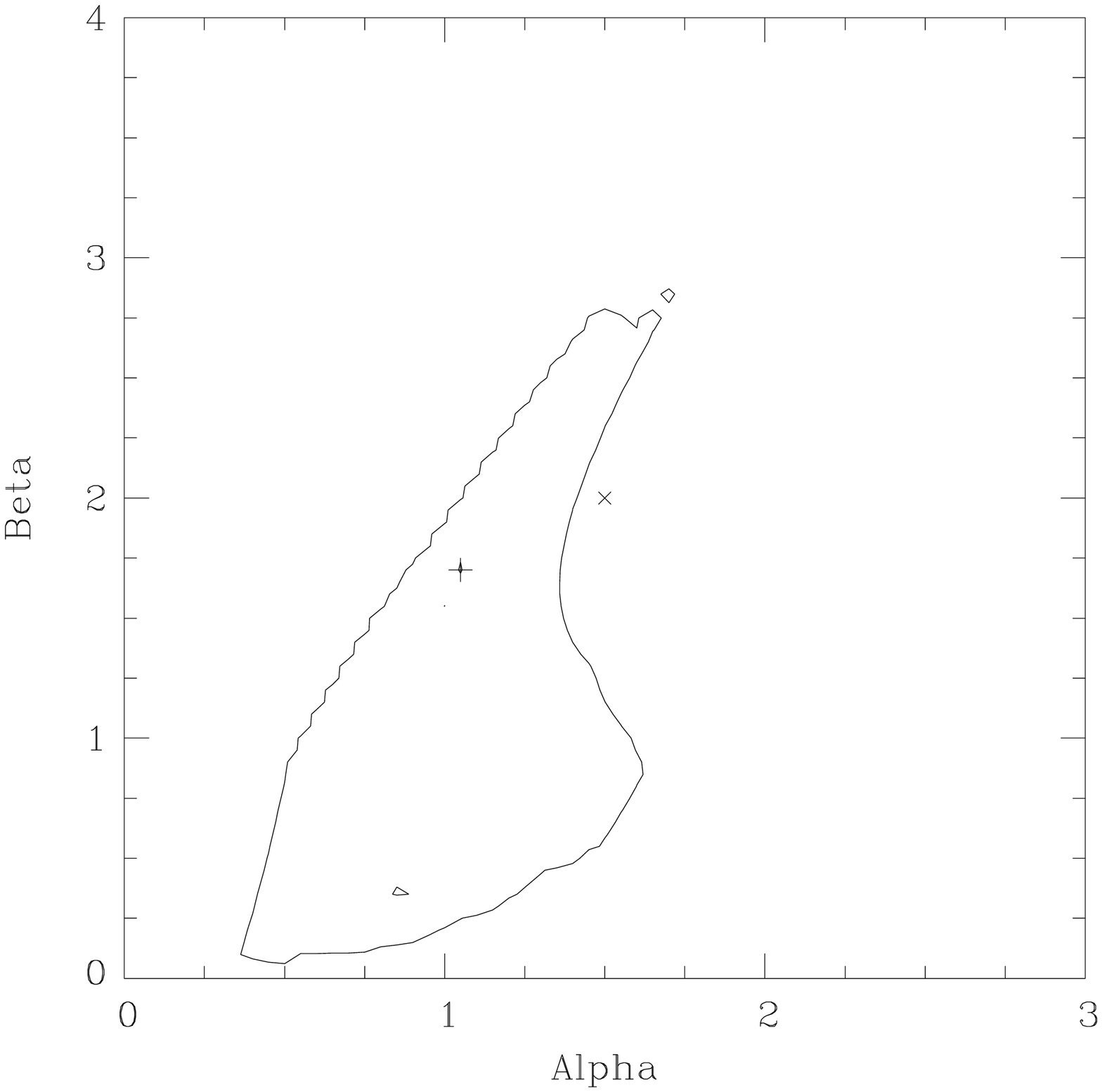,height=6in}

\figcaption[f5.eps]{K-S probability contour map for the
dependencies of hard globular cluster sources
($10^{32}<L_X(0.5-2.5)<10^{33}$ ergs s$^{-1}$) upon $\rho_0$ and $r_c$.
Symbols and contours same as figure 4, except that no 90\% contour
is indicated.  Data are bright hard sources from all clusters in Table
1 except NGC 6366 only.  
}

\psfig{file=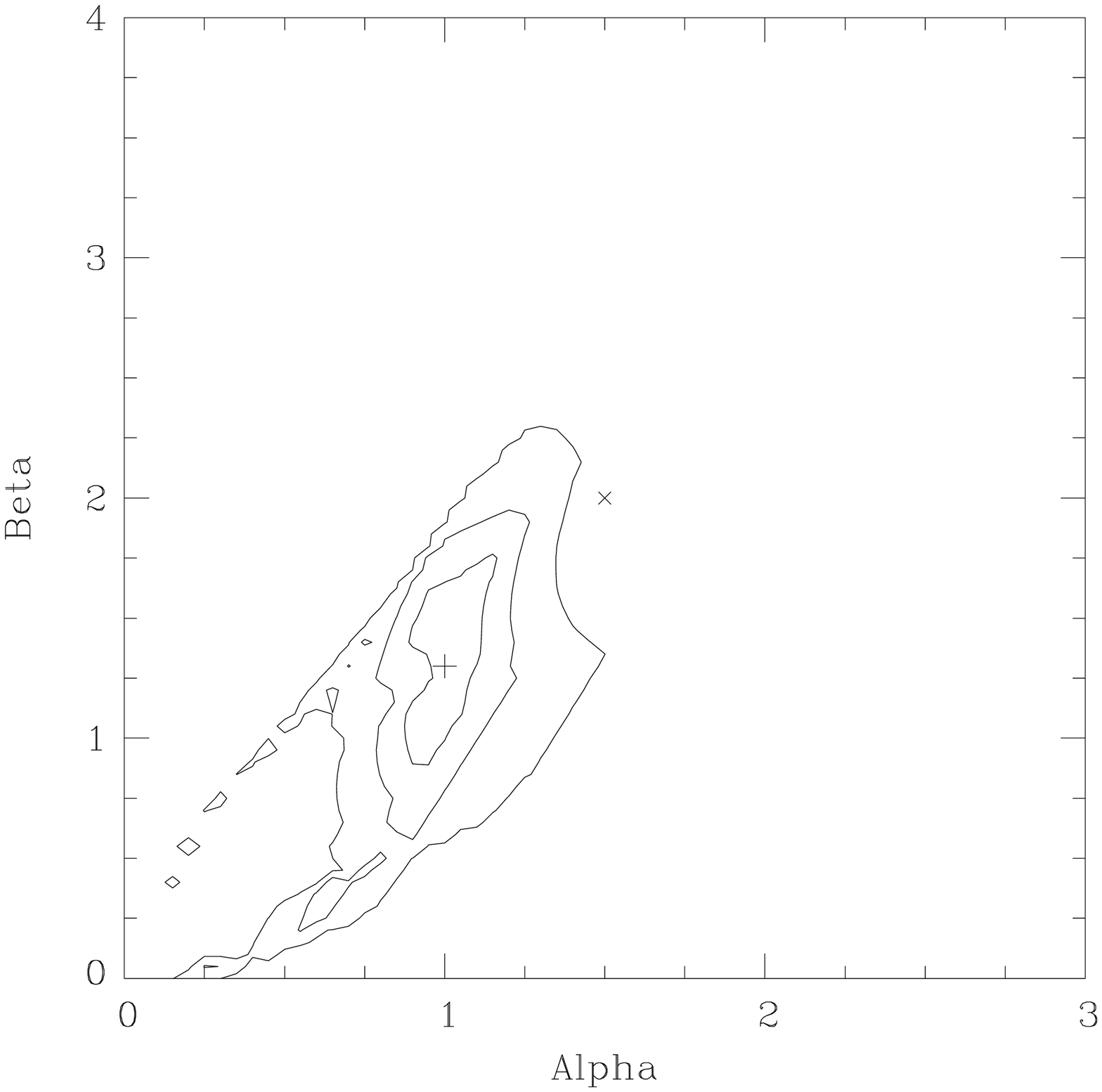,height=6in}

\figcaption[f6.eps]{K-S probability contour map for the
dependencies of hard globular cluster sources
($10^{31}<L_X(0.5-2.5)<10^{33}$ ergs s$^{-1}$) upon $\rho_0$ and $r_c$.
Symbols and contours same as figure 4. Data are from Table 1,
omitting Terzan 5, M13, NGC 6440, and NGC 6366.
}

\psfig{file=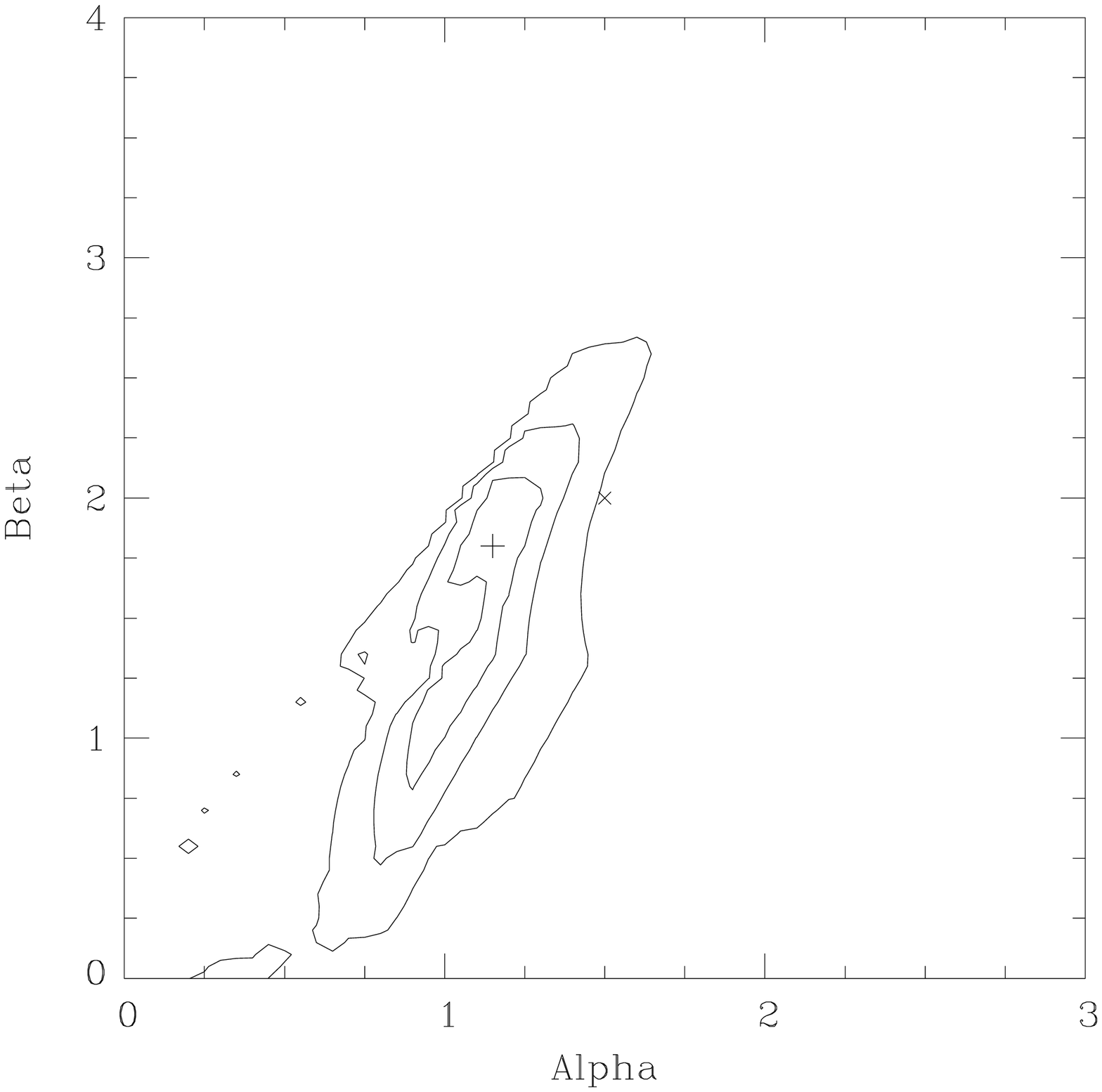,height=6in}

\figcaption[f7.eps]{Same as figure 6, except that NGC
6397 is also excluded from the data.
}

\end{document}